\documentclass[11pt,eqs]{article}
\usepackage{latexsym,amsmath}
\usepackage{hyperref}
\usepackage{ulem}

\def\cleq{\setcounter{equation}{0}}
\title{Generalizations of Courant brackets in the T-dual description of a weakly curved string background}
\author{I. Ivani\v sevi\'c \thanks{e-mail: ivanisevic@ipb.ac.rs}\\
{\it Institute of Physics, University of Belgrade}\\
{\it Pregrevica 118, 11080 Belgrade, Serbia}}

\begin{document}
\maketitle
\begin{abstract}
When T-duality is realized as a canonical transformation, the standard Courant bracket is T-dual invariant, while the $B$- and $\theta$-twisted Courant brackets are exchanged. We determine how these relations are modified for a closed bosonic string moving in a weakly curved background with a constant metric and a Kalb-Ramond field with linear coordinate dependence. We work to first order in its constant field strength. The T-dual background fields depend on a line integral of auxiliary worldsheet gauge fields, making the dual description nonlocal. We construct generalized charges whose parameters depend on this line integral and calculate their Poisson bracket algebra using the T-duality transformation laws and the canonical brackets of the initial theory. The resulting algebra yields a weakly curved generalization of the Courant bracket and its associated Courant algebroid. We establish an explicit isomorphism with the standard Courant algebroid by transporting the bracket, pairing, and anchor. Applying $B$- and $\theta$-shifts to the dual currents, we obtain weakly curved generalizations of the respective twisted Courant brackets. We show that these brackets provide the T-dual realizations of the initial $\theta$- and $B$-twisted Courant brackets, respectively, and identify the corresponding Courant algebroid isomorphisms.

\end{abstract}

\section{Introduction}
\cleq

T-duality plays a central role in string theory, relating theories defined on backgrounds with different geometries and, in some cases, topologies. It forms part of the network of dualities connecting the five superstring theories and M-theory. Buscher developed a systematic procedure for backgrounds admitting a global Abelian isometry \cite{buscher, buscher1}. The procedure gauges this isometry and introduces Lagrange multipliers that enforce vanishing gauge field strength. Integrating out the multipliers recovers the original theory, while integrating out the gauge fields yields its T-dual. Along the T-dualized directions, the canonical momenta of each theory are proportional to the spatial world-sheet derivatives of the coordinates of its T-dual theory. In this setting, T-duality can be understood as a canonical transformation \cite{tdual-can}.

The spatial coordinate derivatives and canonical momenta naturally enter generalized currents whose parameters combine a vector field and a one-form. Their geometric setting is the generalized tangent bundle $TM\oplus T^*M$, the direct sum of the tangent and cotangent bundles of the target manifold $M$. Generalized geometry provides a common framework for studying complex, symplectic, and Poisson structures \cite{gualtieri, calabi-yau}. The charges obtained by integrating these currents generate diffeomorphisms and gauge transformations of the Kalb-Ramond field \cite{doucou, nick1, cdual}. Their Poisson algebra is governed by the Courant bracket, which extends the Lie bracket to sections of the generalized tangent bundle \cite{courant}.

More general twists of the Courant bracket can accommodate both geometric and non-geometric fluxes, conventionally denoted by $H$, $f$, $Q$ and $R$ \cite{flux1, flux2, flux3}. Such structures have been studied in the framework of proto-Lie bialgebroids \cite{royt, royt1}, through combined two-form and bivector twists \cite{CBTh, ljubica}, and using graded symplectic methods \cite{heller}. Broader bialgebroid constructions have also been investigated \cite{keremcan}, while generalized fluxes can be encoded in deformations of string current algebras \cite{osten}.

A particularly useful geometric interpretation of T-duality arises in topological T-duality. For T-dual principal torus bundles over a common base, equipped with compatible closed three-form fluxes, T-duality induces an isomorphism between the corresponding reduced Courant algebroids over that base \cite{Tdualiso}. A related isomorphism connects the Courant algebroid structures associated with the $B$-twisted and $\theta$-twisted Courant brackets appearing in the Poisson algebras of generalized charges \cite{c, crdual}. The T-duality interpretation of these current algebra results uses the standard exchange of canonical momenta and spatial coordinate derivatives. This raises the question of how the current algebras and associated algebroid structures are modified when the T-duality relations acquire background-dependent corrections.

A useful setting for addressing this question is the weakly curved background, characterized by a constant metric tensor and a Kalb-Ramond field with a linear coordinate-dependent part proportional to a constant, infinitesimal field strength \cite{wcb, wcb3, wcb4, danijel1, danijel2}. Throughout, we retain terms only to first order in this field strength. The action remains invariant under the global shift symmetry of the coordinates; however, it does not become covariant merely by replacing partial derivatives with covariant ones. As such, the Buscher procedure must be modified by introducing a gauge-invariant coordinate $\Delta x^\mu$, defined as a line integral of the covariant derivatives of the original coordinates. The resulting T-dual background fields depend on both the dual coordinates $y_\mu$ and their doubles $\tilde y_\mu$, defined through $\Delta\tilde y_\mu=\int_P(d\tau y^{\prime}_{\mu}+d\sigma \dot{y}_\mu)$. Unlike the integral of $dy_\mu$, this expression cannot generally be reduced to an endpoint difference of $y_\mu$. Consequently, the background fields are nonlocal functionals of $y_\mu$, making the T-dual theory nonlocal in these variables. With the Poisson structure induced from the original theory, the T-dual phase space variables therefore do not obey the standard canonical brackets; in particular, the T-dual coordinates are non-commutative \cite{NC0, NC, NCwcb}.

In this paper, we investigate the Courant algebroid structures arising in the T-dual description of a closed bosonic string in a weakly curved background. We construct generalized currents from the spatial derivatives of the T-dual coordinates and the T-dual momenta. To reflect the nonlocality of the dual theory, we choose their parameters to depend on the line integral variable entering the dual background fields. Using the T-duality relations, we express these currents in terms of the original phase space variables and compute their Poisson brackets. The resulting algebra of generalized charges defines a weakly curved generalization of the Courant bracket. We establish an isomorphism with the standard Courant algebroid and determine how the bracket, anchor, and symmetric pairing are transported. In particular, the symmetric pairing in the chosen T-dual basis differs from the standard $O(D,D)$ pairing by corrections inherited from the modified T-duality relations.

We also construct generalized currents in the $B$ and $\theta$ frames and derive the corresponding twisted versions of the weakly curved Courant bracket. For each frame, we establish an isomorphism with the corresponding standard twisted Courant algebroid and determine how it relates the brackets, anchors, and symmetric pairings.

The paper is organized as follows. Section 2 reviews the Courant algebroid structures of the initial theory, introduces the weakly curved string background and presents the generalized Buscher procedure in the Lagrangian and Hamiltonian formulations. Section 3 constructs generalized currents and charges in the T-dual phase space basis, derives their Poisson algebra and the resulting weakly curved Courant bracket, and establishes the isomorphism with the standard Courant algebroid. Sections 4 and 5 develop the analogous constructions in the $B$ and $\theta$ frames, respectively.

\section{Model: weakly curved background}
\cleq

We consider a closed bosonic string whose classical dynamics are governed by the $\sigma$-model action
\begin{equation}\label{eq:action}
S= \kappa \int_{\Sigma} d^2\xi\Big[\frac{1}{2}\eta^{\alpha\beta}G_{\mu\nu}[x]
+\varepsilon^{\alpha\beta}B_{\mu\nu}[x]\Big]
\partial_{\alpha}x^{\mu}\partial_{\beta}x^{\nu} \, ,
\end{equation}
where $\Sigma$ is the world-sheet parametrized by $\xi^{\alpha}$ (the time-like parameter $\xi^0 = \tau$ and space-like parameter $\xi^1 = \sigma$), and $\eta^{\alpha \beta}$ is the intrinsic worldsheet metric. The background fields that describe the motion of the string are the metric tensor $G_{\mu \nu}$ and the antisymmetric Kalb-Ramond field $B_{\mu \nu}$, while the dilaton $\Phi$ can be omitted, since we are working in classical theory. The indices $\mu, \nu$ denote the $D$-dimensional space-time, $\varepsilon^{\alpha \beta}$ is the Levi-Civita symbol ($\varepsilon^{01}= -1$), and $\kappa = \frac{1}{2 \pi \alpha^{\prime}}$, where $\alpha^{\prime}$ is the Regge slope parameter. 

To the lowest order in slope parameter $\alpha^\prime$, quantum conformal invariance requires the background fields to satisfy the following equations of motion:
\begin{eqnarray} \label{eq:eom-conf}
R_{\mu \nu} - \frac{1}{4} B_{\mu \rho \sigma}
B_{\nu}^{\ \rho \sigma}=0,\quad
D_\rho B^{\rho}_{\ \mu \nu} = 0,
\end{eqnarray}
where
\begin{eqnarray} \label{eq:Hdef}
B_{\mu\nu\rho}=\partial_\mu B_{\nu\rho}
+\partial_\nu B_{\rho\mu}+\partial_\rho B_{\mu\nu}
\end{eqnarray}
is the field strength of $B_{\mu \nu}$, and
$R_{\mu \nu}$ and $D_\mu$ are the Ricci tensor and the
covariant derivative with respect to the space-time metric. 

The weakly curved background \cite{wcb} is a well-known non-trivial solution of equations (\ref{eq:eom-conf}), defined by
\begin{eqnarray}\label{eq:wcb}
G_{\mu\nu}=\text{const}, \quad B_{\mu\nu}[x]=b_{\mu\nu}+
\frac{1}{3}B_{\mu\nu\rho}x^\rho
\equiv 
b_{\mu\nu}+
h_{\mu\nu}[x] \, ,
\end{eqnarray}
where the Kalb-Ramond field strength $B_{\mu \nu \rho}$ (\ref{eq:Hdef}) is taken to be constant and infinitesimal - all terms in the second and  higher order in $B_{\mu \nu \rho}$ are neglected in future calculations.

The canonical Hamiltonian of the initial theory is given by 
\begin{equation} \label{eq:Hcdef}
{\cal H}_C = \pi_{\mu} \dot{x}^\mu - {\cal L} = \frac{1}{2\kappa} \pi_\mu (G^{-1})^{\mu \nu}\pi_\nu + \frac{\kappa}{2} x^{\prime \mu} G^E_{\mu \nu} x^{\prime \nu}- 2  x^{\prime \mu} B_{\mu \rho} (G^{-1})^{\rho \nu}  \pi_\nu \, ,
\end{equation}
where $\pi_\mu$ are the canonical momenta, given by
\begin{equation}
\pi_{\mu} = \frac{\partial {\cal L}}{\partial \dot{x}^\mu} = \kappa G_{\mu \nu} \dot{x}^\nu - 2 \kappa B_{\mu \nu} x^{\prime \nu} \, ,
\end{equation}
and $G_E$ is the effective metric, given by 
\begin{equation} \label{eq:Geff}
G^E_{\mu \nu} = G_{\mu \nu} - 4 (B G^{-1} B)_{\mu \nu} \, .
\end{equation}
The standard Poisson bracket relations are assumed
\begin{equation} \label{eq:PB-def}
\{ x^\mu (\sigma), x^\nu (\bar{\sigma}) \} = 0 \, , \quad \{ \pi_\mu (\sigma), \pi_\nu (\bar{\sigma}) \} = 0 , \quad \{ x^\mu (\sigma), \pi_\nu (\bar{\sigma}) \} = \delta^\mu_\nu \delta (\sigma-\bar{\sigma}) \, . 
\end{equation}

\subsection{Courant algebroids in the initial theory}

In the matrix notation, the canonical Hamiltonian can be expressed as
\begin{eqnarray} \label{eq:Hcmatrixdef}
{\cal  H}_C = \frac{1}{2\kappa} (X^T)^M H_{MN} X^N \, ,
\end{eqnarray}
where $H_{MN}$ is the generalized metric, given by
\begin{equation} \label{eq:genmet}
H_{MN} = \begin{pmatrix}
G^E_{\mu \nu} & -2 (BG^{-1})_{\!\mu}^{\ \nu} \\
2 (G^{-1}B)^\mu_{\ \nu} & (G^{-1})^{\mu \nu} 
\end{pmatrix}\, ,
\end{equation}
and $X^M$ is double canonical variable, defined by
\begin{equation} \label{eq:XM-def}
X^M = \begin{pmatrix}
\kappa x^{\prime \mu} \\
\pi_\mu
\end{pmatrix} \, . 
\end{equation}
The index $M$ can take values from 0 to $2D-1$ and includes both covariant and contravariant components. Since the double canonical variable includes vector and one-form components, it has the structure of the generalized vector. The generalized vectors are sections of the generalized tangent bundle, which is just a direct sum of tangent and cotangent bundle over a manifold. The geometry of the generalized tangent bundle is known as generalized geometry \cite{gualtieri, calabi-yau}.
There is a natural symmetric pairing between two generalized vectors, defined by 
\begin{equation} \label{eq:eta-def}
\langle \Lambda_1, \Lambda_2 \rangle = \Lambda_1^M \eta_{MN} \Lambda_2^N \, , \quad \eta_{MN}  = \begin{pmatrix}
0 & 1 \\
1 & 0 
\end{pmatrix} \, .
\end{equation}
The Courant bracket \cite{courant} is a natural generalization of the Lie bracket, which acting on two generalized vector results in a generalized vector $[\Lambda_1, \Lambda_2]_{\cal C} = \Lambda = \xi \oplus \lambda$ with components given by 
\begin{equation} \label{eq:courant-nc}
\xi = [\xi_1, \xi_2]_L \, , \quad \lambda = {\cal{L}}_{\xi_1} \lambda_2 - {\cal{L}}_{\xi_2} \lambda_1 - \frac{1}{2}d (i_{\xi_1} \lambda_2 - i_{\xi_2} \lambda_1) \, , 
\end{equation}
where standard notation for Lie derivative, interior product, and exterior derivative are assumed. It can be easily obtained by considering the so called generalized charges in the form 
\begin{equation} \label{eq:G-def}
{\cal Q}_{\Lambda} = \int_0^{2\pi} d\sigma \Lambda^M \eta_{MN} X^N \, , \quad \Lambda^M = \begin{pmatrix}
\xi^\mu \\
\lambda_\mu
\end{pmatrix} \, ,
\end{equation}
and then calculating the standard Poisson bracket between two sets of charges \cite{c, nick1, cdual}
\begin{equation} \label{eq:GG-courant}
\{ {\cal Q}_{\Lambda_1},{\cal Q}_{\Lambda_2} \}  = - {\cal Q}_{[\Lambda_1, \Lambda_2]_{\cal C}} \, .
\end{equation}
An important property of the charges (\ref{eq:G-def}) is that they also generate general coordinate transformations and local gauge transformations \cite{dualsim}. The Courant bracket also appears from Siegel construction in double field theory \cite{siegel, siegel1} by projecting out half of the coordinates \cite{doucou, twistedC}.

Together with Courant bracket, the generalized tangent bundle is equipped with a Courant algebroid \cite{courant1}, a vector bundle equipped with a skew-symmetric bracket on its smooth sections, a nondegenerate symmetric pairing, and an anchor map to the tangent bundle, satisfying appropriate compatibility conditions. The complete definition and the conventions used here are recalled in Appendix [\ref{app:courant}]. The standard Courant algebroid $(TM\oplus T^*M, [, ]_{{\cal C}}, \eta, \rho_0 )$ consists of the generalized tangent bundle as its vector bundle, the Courant bracket (\ref{eq:courant-nc}), the $O(D, D)$ invariant pairing $\eta$ (\ref{eq:eta-def}), and the projection to the tangent bundle $\rho_0(\xi \oplus \lambda) = \xi$ as its anchor.

The Hamiltonian can be expressed in a different basis from the canonical one (\ref{eq:XM-def}), in the diagonal quadratic form. These bases are known as noncanonical currents, and are characterized by the emergence of fluxes in their Poisson bracket relations \cite{flux2, flux3}. In one case, we can transform the generalized metric by 
\begin{eqnarray} \label{eq:H-B}
&& H_{MN} = ( (e^{\hat{B}})^T)_M^{\ K}\ G_{KL}\ (e^{\hat{B}})^{\!L}_{\ N} \, , \\ \notag
&& G_{MN} = 
\begin{pmatrix}
G_{\mu \nu} & 0 \\
0 & (G^{-1})^{\mu \nu}
\end{pmatrix} \, , \quad (e^{\hat{B}})^{\!M}_{\ N} = \begin{pmatrix}
\delta^\mu_\nu & 0 \\
2 B_{\mu \nu} & \delta^\nu_\mu
\end{pmatrix} \, ,
\end{eqnarray}
so that the canonical Hamiltonian is given by
\begin{equation}
{\cal H}_{\cal C} = \frac{1}{2\kappa} X_{\hat B}^M G_{MN} X_{\hat B}^N \, , \quad X_{\hat B}^M = (e^{\hat{B}})^{\!M}_{\ N}\ X^N = 
\begin{pmatrix}
\kappa x^{\prime \mu} \\
\pi_\mu + 2 \kappa B_{\mu \nu} x^{\prime \nu}  
\end{pmatrix} \equiv
\begin{pmatrix}
\kappa x^{\prime \mu} \\
\iota_\mu
\end{pmatrix}
\, .
\end{equation}
The generalized charges (\ref{eq:G-def}) expressed in this noncanonical basis have the form
\begin{eqnarray} \label{eq:GBdef}
{\cal Q}^{\hat{B}}_{\Lambda}= \int d\sigma (\Lambda^T)^M \eta_{MN} X_{\hat B}^N  = \int d\sigma \Big( \xi^\mu \iota_\mu + \lambda_\mu \kappa x^{\prime \mu} \Big)  \, ,
\end{eqnarray}
and its algebra closes on the $B$-twisted Courant bracket
\begin{equation} \label{eq:GBGB-GB}
\Big\{ {\cal{Q}}^{\hat{B}}_{\Lambda_1},  {\cal{Q}}^{\hat{B}}_{\Lambda_2} \Big\} = - {\cal{Q}}^{\hat{B}}_{[\Lambda_1, \Lambda_2]_{{\cal C}_{B}}} \, , \quad [\Lambda_1, \Lambda_2]_{{\cal C}_B} = e^{\hat{B}}[e^{-\hat{B}}\Lambda_1, e^{-\hat{B}}\Lambda_2]_{\cal C}  \, .
\end{equation}
Assuming the following notation $[\Lambda_1, \Lambda_2]_{{\cal C}_B} = \Lambda = \xi \oplus \lambda$, the components of the result of the $B$-twisted Courant bracket on two generalized vectors are given by
\begin{equation} \label{eq:Bcourant-nc}
\xi = [\xi_1, \xi_2]_L \, , \quad \lambda = {\cal{L}}_{\xi_1} \lambda_2 - {\cal{L}}_{\xi_2} \lambda_1 - \frac{1}{2}d (i_{\xi_1} \lambda_2 - i_{\xi_2} \lambda_1) + 2d B (\xi_1, \xi_2, .)\, , 
\end{equation}
where the additional term $2d B (\xi_1, \xi_2, .)$ represents the contraction of two vector fields $\xi_1$ and $\xi_2$ with flux term $dB$. The corresponding Courant algebroid is $(TM\oplus T^*M, [, ]_{{\cal C}_B}, \eta, \rho_0 )$.

In another case, we can express the generalized metric in $\theta$-frame as
\begin{eqnarray}  \label{eq:H-th}
&& H_{MN} = ((e^{\hat{\theta}})^T)_M^{\ K}\ {}^\star G_{KL} (e^{\hat{\theta}})^{\!L}_{\ N} \, ,  \\ \notag
&& {}^\star G_{MN} = 
\begin{pmatrix}
G^E_{\mu \nu} & 0 \\
0 & (G_E^{-1})^{\mu \nu}
\end{pmatrix} \, , \quad 
(e^{\hat{\theta}})^{\!M}_{\ N} = \begin{pmatrix}
\delta^\mu_\nu & \kappa \theta^{\mu \nu} \\ 
0 & \delta^\nu_\mu
\end{pmatrix} \, ,
\end{eqnarray}
where $\theta^{\mu \nu}$ is the non-commutativity parameter, given by
\begin{equation} \label{eq:theta-def}
\theta^{\mu\nu}=
-\frac{2}{\kappa}
(G^{-1}_{E}BG^{-1})^{\mu\nu} \, .
\end{equation}
The non-commutativity parameter, which has the structure of a bi-vector, is seen as T-dual to the Kalb-Ramond $B$ field. The Hamiltonian can now be expressed by
\begin{equation}
{\cal \hat{H}}_C = \frac{1}{2 \kappa}  X_{\hat{\theta}}^M {^\star G}_{MN} X_{\hat{\theta}}^N\, , \quad X_{\hat{\theta}}^M = (e^{\hat{\theta}})^{\!M}_{\ N}\ X^N =  
\begin{pmatrix}
\kappa x^{\prime \mu} + \kappa \theta^{\mu \nu} \pi_\nu \\
\pi_\mu
\end{pmatrix} \equiv
\begin{pmatrix}
k^\mu \\
\pi_\mu 
\end{pmatrix} \, .
\end{equation}
Now the generalized charges (\ref{eq:G-def}) become
\begin{eqnarray} \label{eq:Gthdef}
{\cal Q}^{\hat{\theta}}_{\Lambda}=  \int d\sigma (\Lambda^T)^M \eta_{MN} X_{\hat{\theta}}^N = \int d\sigma \Big( \xi^\mu \pi_\mu + \lambda_\mu k^{\mu} \Big)  \, ,
\end{eqnarray}
which gives rise to the $\theta$-twisted Courant bracket
\begin{equation} \label{eq:Gth-Gth}
\Big\{ {\cal Q}^{\hat{\theta}}_{\Lambda_1}, \, {\cal Q}^{\hat{\theta}}_{\Lambda_2} \Big\} = -{\cal Q}^{\hat{\theta}}_{[\Lambda_1, \Lambda_2]_{{\cal C}_{\theta}}} \, , \quad [\Lambda_1, \Lambda_2]_{{\cal C}_{\theta}}  = e^{\hat{\theta}}[e^{-\hat{\theta}}\Lambda_1, e^{-\hat{\theta}}\Lambda_2]_{\cal C} \, .
\end{equation}
Explicitly, this bracket yields a vector and one-form components $[\Lambda_1, \Lambda_2]_{{\cal C}_{\theta}} = \Lambda = \xi \oplus \lambda$, given by
\begin{align} \label{eq:thcourant-nc}
\xi = &  [\xi_1,\xi_2]_L - \kappa [\xi_2, \lambda_1 \theta]_L + \kappa [\xi_1,\lambda_2 \theta]_L + \frac{\kappa^2}{2} [\theta,\theta]_S (\lambda_1, \lambda_2, .) \\ \notag
& -\kappa \theta \Big( {\cal{L}}_{\xi_1} \lambda_2 -{\cal{L}}_{\xi_2} \lambda_1 - \frac{1}{2} d (i_{\xi_1} \lambda_2 - i_{\xi_2} \lambda_1 ) \Big) \\ \notag
\lambda = & {\cal{L}}_{\xi_1}\lambda_2 - {\cal{L}}_{\xi_2} \lambda_1 -\frac{1}{2} d (i_{\xi_1} \lambda_2 - i_{\xi_2} \lambda_1 ) + \kappa [\lambda_1 ,\lambda_2]_{ \theta} \, , \notag
\end{align}
where $ [\theta,\theta]_S$ represents the Schouten-Nijenhuis bracket \cite{SNB}, which is the Lie bracket extension to multi-vectors, in this context given by
\begin{equation}
\left. [\theta, \theta]_S \right| ^{\mu \nu \rho} = 2 (\theta^{\mu \sigma}\partial_{\sigma}\theta^{\nu \rho} + \theta^{\nu \sigma}\partial_{\sigma}\theta^{\rho \mu} +\theta^{\rho \sigma}\partial_{\sigma}\theta^{\mu \nu }) \, ,
\end{equation}
and $[\lambda_1, \lambda_2]_{\theta}$ is the Koszul bracket \cite{koszul}, given by
\begin{equation} \label{eq:koszul-def}
[\lambda_1, \lambda_2]_{\theta} = {\cal L}_{\theta (\lambda_1)}\lambda_2-{\cal L}_{\theta (\lambda_2)}\lambda_1-d (\theta(\lambda_1, \lambda_2)) \, ,
\end{equation}
and $\theta(.)$ is defined 
\begin{equation} \label{eq:theta(.)}
\theta(\lambda_1) \lambda_2 = \theta(\lambda_1, \lambda_2) \, , \quad \Big( \theta(\lambda_1)\Big)^{\! \mu} = \lambda_{1\nu} \theta^{\nu \mu} \, .
\end{equation}
The corresponding Courant algebroid is $(TM\oplus T^*M, [, ]_{{\cal C}_{\theta}}, \eta, \rho_{\theta} = \rho_0 \circ e^{-\hat{\theta}} )$ \cite{CA-Dirac}.

In the standard canonical realization of T-duality \cite{tdual-can, tdual-can1}, the simultaneous exchanges of variables with their T-dual analogs 
\begin{equation} \label{eq:selfTdual}
\pi_\mu \leftrightarrow \kappa x^{\prime\mu},
\qquad
2B_{\mu\nu} \leftrightarrow \kappa\theta^{\mu\nu}
\end{equation}
map the noncanonical bases used to express the $B$-frame and $\theta$-frame generalized charges, ${\cal Q}^{\hat B}_{\Lambda}$ (\ref{eq:GBdef})  and ${\cal Q}^{\hat\theta}_{\Lambda}$ (\ref{eq:Gthdef}), into one another. The Courant algebroid isomorphism (\ref{eq:CA-isom}) between these structures is given by \cite{crdual}
\begin{equation}
\varphi  [\Lambda_1, \Lambda_2]_{{\cal C}_B} =  \Big( [ \varphi (\Lambda_1), \varphi (\Lambda_2)]_{{\cal C}_{\theta}} \Big) \, , \quad \varphi = e^{\hat{\theta}}e^{-\hat{B}} \, .
\end{equation}
In subsequent sections, we will investigate how these relations are modified in a weakly curved background, where T-duality cannot be seen as canonical transformation.

\subsection{Weakly curved background - from initial to T-dual theory}

The weakly curved background is interesting because it is a non-trivial theory where an adapted version of the Buscher procedure can be applied and T-dual theory is known. We sketch the procedure here and for more computational details the reader is referred to \cite{wcb}. We begin by rewriting the action (\ref{eq:action}) in terms of the light-cone coordinates:
\begin{equation} \label{eq:action1}
S = \kappa \int_{\Sigma} d^2\xi\, \partial_+ x^\mu \Pi_{+ \mu \nu}[x] \partial_- x^\nu \, , 
\end{equation}
where 
\begin{equation} \label{eq:Pi-pm}
\Pi_{\pm\mu\nu}[x]= B_{\mu\nu}[x]\pm\frac{1}{2}G_{\mu\nu}[x] \, ,
\end{equation}
and
\begin{equation} \label{eq:lightcone}
\xi^{\pm} = \frac{1}{2}(\tau \pm \sigma),\quad \partial_\pm = \partial_\tau \pm \partial_\sigma \, .
\end{equation}
The action remains invariant under the global shift symmetry of the coordinates $\delta x^\mu = \lambda^\mu$. In the case of constant backgrounds, the global shift symmetry is localized by introducing the independent world-sheet gauge fields $v_\alpha^\mu$ and substituting the partial derivatives with the covariant ones
\begin{equation}
\partial_{\alpha}x^\mu\rightarrow
D_{\alpha}x^\mu=\partial_\alpha x^\mu
+v^{\mu}_{\alpha} \, ,
\end{equation}
where $\delta v^{\mu}_{\alpha}=-\partial_\alpha \lambda^\mu$ ensures the gauge invariance. In the case of weakly curved background, this is not enough, since the background field $B_{\mu\nu}$ depends on the coordinate $x^\mu$, which is not gauge invariant. Therefore, the additional step has to be implemented, where the coordinate is replaced by some extension of it. Let us define the invariant coordinate by
\begin{eqnarray}
\Delta x^{\mu}_{inv}&\equiv&
\int_{P}d\xi^\alpha\,D_\alpha x^\mu
=\int_{P}(d\xi^{+}D_{+}x^{\mu}
+d\xi^{-}D_{-}x^{\mu})
\nonumber\\
&=&
x^\mu-x^\mu(\xi_{0})
+\Delta V^\mu,
\end{eqnarray}
where
\begin{equation}\label{eq:vdef}
\Delta V^\mu
\equiv
\int_{P}d\xi^\alpha v^{\mu}_{\alpha}
=\int_{P}(d\xi^{+} v^{\mu}_{+}
+d\xi^{-} v^{\mu}_{-}).
\end{equation}
The line integral is taken along the path $P$, from the initial point $\xi^{\alpha}_{0}(\tau_{0},\sigma_{0})$ to the final one $\xi^{\alpha}(\tau,\sigma)$. Locally, the integral is independent of the path once the vanishing field strength constraint imposed by the Lagrange multiplier is satisfied \cite{wcb}. 

From now on, the T-dualization procedure follows the same steps as the standard Buscher procedure. The gauge-invariant action is given by 
\begin{eqnarray}\label{eq:ainv}
S_{inv}&=&\kappa\int d^{2}\xi\Big{[}
D_{+}x^\mu \Pi_{+\mu\nu}[\Delta x_{inv}] D_{-}x^\nu
+\frac{1}{2} y_\mu (\partial_{+}v^{\mu}_{-} - \partial_{-} v^{\mu}_{+})
\Big{]} \, , 
\end{eqnarray}
where $y_\mu$ is the Lagrange multiplier enforcing the vanishing of the gauge-field strength. Furthermore, we can fix the gauge by imposing $x^\mu(\xi)=x^\mu(\xi_{0})$. The equations of motion are obtained by varying with respect to the Lagrange multiplier $y_\mu$ and the gauge fields $v^\mu_\pm$. Solving the former equations of motion returns the initial theory, while solving the latter yields the T-dual theory (see \cite{wcb} for a detailed derivation).

The T-dualization rules are given by
\begin{equation}\label{eq:xcong}
\partial_{\pm}x^\mu\cong
-\kappa\Theta^{\mu\nu}_{\pm}[\Delta V]
\Big{[}
\partial_{\pm} y_\nu
\pm 2\beta^{\mp}_{\nu} [V]
\Big{]},
\end{equation}
where 
\begin{equation}\label{eq:beta}
\beta^\pm_\mu[x]=\frac{1}{2}(\beta^0_\mu\pm\beta^1_\mu)=
\mp\frac{1}{2}
h_{\mu\nu}[x]\partial_{\mp} x^\nu,\,
\beta^0_\mu[x]=h_{\mu\nu}[x]x'^\nu,\,
\beta^1_\mu[x]=-h_{\mu\nu}[x]\dot x^\nu\, ,
\end{equation}
and
\begin{eqnarray}\label{eq:Theta-def}
{\Theta}^{\mu\nu}_{\pm} =\theta^{\mu\nu} \mp\frac{1}{\kappa}(G_{E}^{-1})^{\mu\nu} \, .
\end{eqnarray}
It is not difficult to show that these fields satisfy following decomposition of identity
\begin{equation} \label{eq:decomp-fields}
(G G_E^{-1})_\mu^{\ \nu} + 2\kappa (B\theta)_\mu^{\ \nu} = \delta_\mu^\nu
\end{equation}
We denote the constant parts of the effective metric and the noncommutativity parameter by $g_E$ and $\theta_0$, respectively:
\begin{equation}
g^E_{\mu\nu}
=G_{\mu\nu}-4(bG^{-1}b)_{\mu\nu},
\qquad
\theta_0^{\mu\nu}
=-\frac{2}{\kappa}(g_E^{-1}bG^{-1})^{\mu\nu}.
\end{equation}

The application of (\ref{eq:xcong}) to (\ref{eq:action1}) gives the T-dual action
\begin{equation}\label{eq:action-dual}
^\star S[y]=
\frac{\kappa^{2}}{2}
\int d^{2}\xi\
\partial_{+}y_\mu
\Theta_{-}^{\mu\nu}[\Delta V^{(0)}(y)]
\partial_{-}y_\nu \, ,
\end{equation}
from which we have
\begin{eqnarray}\label{eq:Tdual-fields}
&G_{\mu\nu}&\rightarrow\,
{^\star G}^{\mu\nu}=
(G^{-1}_{E})^{\mu\nu}[\Delta V^{(0)}
] \, ,
\nonumber\\
&B_{\mu\nu}[x]&\rightarrow\,
{^{\star}B}^{\mu\nu}=
\frac{\kappa}{2}\theta^{\mu\nu}[\Delta V^{(0)}
] \, .
\end{eqnarray}
The resulting background fields depend on the line integral of the world-sheet fields \(v_\alpha^\mu\) along a path \(P\) from a fixed reference point \(\xi_0\) to \(\xi\). 
The line integral is retained through first order in the Kalb-Ramond field strength, $V^\mu=V_0^\mu +V_1^\mu$. The correction $V_1^\mu $ is obtained from the first-order solution for the gauge fields; its explicit expression is given in Appendix \ref{app:v}. All subsequent parameter dependence on $V^\mu$ includes this first-order contribution.

The zeroth-order contribution is
\begin{equation} 
 V_0^\mu(y)
=
-\kappa\theta_0^{\mu\nu} y_\nu^{(0)}
+
\left(g_E^{-1}\right)^{\mu\nu}
\widetilde{y}_\nu^{(0)}.
\label{eq:V0}
\end{equation}
where
\begin{equation}\label{eq:tildey}
{\tilde{y}}_\mu=\int_{P}(d\tau y^\prime_\mu+d\sigma \dot{y}_\mu) \, ,
\end{equation}
while $\theta_0$ and $g^{-1}_E$ are the zeroth-order terms of the non-commutativity parameter $\theta$ and the effective metric inverse $G_E^{-1}$, respectively.

The argument of the background fields in the T-dual action is a line integral that depends on two sets of coordinates related by 
\begin{equation} \label{eq:y-ytilde}
\dot{{\tilde{y}}}_\mu=
{y}_\mu^{\prime},\quad
\tilde{y}_\mu^{\prime}=\dot{y}_\mu \, ,
\end{equation}
and therefore the background does not correspond to a standard manifold, even locally.  Up to the first order, the relation (\ref{eq:xcong}) can be rewritten as \cite{wcb, T-dual-2nd}
\begin{equation}
V^\mu  \cong x^\mu \, .
\label{eq:x-cong-v}
\end{equation}
For completeness, we also write down the inverse T-dualization rules
\begin{equation}\label{eq:ycong}
\partial_{\pm}y_\mu\cong
-2\Pi_{\mp\mu\nu}[\Delta x]\partial_{\pm}x^\nu\mp
2\beta^{\mp}_{\mu}[x] \, ,
\end{equation}
which returns the T-dual action to the initial one.

\subsection{Hamiltonian formulation of T-duality}

Let us begin by obtaining the T-dual Hamiltonian from the T-dual Lagrangian. The T-dual momenta can be obtained by varying the T-dual Lagrangian by both $\dot{y}_\mu$ and $\tilde{y}^{\prime}_{\mu}$, due to relation (\ref{eq:y-ytilde})
We obtain 
\begin{equation} \label{eq:pistar-def}
^\star\pi^\mu=
\kappa(G^{-1}_{E})^{\mu\nu}\big[\Delta  V[y]\big]
\dot{y}_\nu
-\kappa^{2}\theta^{\mu\nu}\big[\Delta  V[y]\big]
y^\prime_\nu
-\kappa  (g_E^{-1})^{\mu\nu}\beta^1_\nu\big[V[y]\big].
\end{equation}
From that, we can express the time-like derivative of the T-dual coordinate
\begin{equation} \label{eq:doty-pi}
\dot{y}_\mu = \frac{1}{\kappa} G^E_{\mu \nu} \big[\Delta  V[y]\big] {}^{\star} \pi^\nu + \kappa G^E_{\mu \rho} \big[\Delta  V[y]\big] \theta^{\rho \nu} \big[\Delta  V[y]\big] y^{\prime}_\nu + \beta_\mu^{1} [V(y)] \, .
\end{equation}

The T-dual Hamiltonian can be obtained by straight-forward calculation, by substituting (\ref{eq:doty-pi}) into ${}^\star {\cal H}_C = {}^\star \pi^\mu \dot{y}_\mu - {}^{\star} {\cal L}$. We obtain
\begin{eqnarray}
{}^\star {\cal H}_C = \frac{1}{2\kappa} {}^\star \pi^\mu G^E_{\mu \nu} {}^\star \pi^\nu + \kappa {}^\star \pi^\mu G^E_{\mu  \rho} \theta^{\rho \nu} y^{\prime}_\nu + \frac{\kappa}{2} y^{\prime}_\mu \Big( (G_E^{-1})^{\mu \nu} - \kappa^2 (\theta G_E \theta)^{\mu \nu} \Big) y^{\prime}_\nu \, .
\end{eqnarray}
For future conveniences, we rewrite the canonical Hamiltonian in the matrix notation, which is more suitable for generalized geometry description
\begin{equation} \label{eq:Hstar-matrix}
{}^\star {\cal H}_C = \frac{1}{2\kappa} {}^{\star} X^M \ {}^\star H_{MN}\big[\Delta  V[y]\big] \ {}^{\star} X^N \, , 
\end{equation}
where ${}^\star X^M$ is T-dual double canonical variable, given by 
\begin{equation} \label{eq:Xstar-def}
{}^\star X^M = \begin{pmatrix}
{}^\star \pi^\mu \\
\kappa y^{\prime}_\mu
\end{pmatrix} \, ,
\end{equation}
and ${}^\star H_{MN}$ is T-dual generalized metric, given by
\begin{equation}
{}^\star H_{MN} = \begin{pmatrix}
G^E_{\mu \nu} & \kappa G^E_{\mu \rho}  \theta^{\rho \nu} \\ 
-\kappa \theta^{\mu \rho} G^E_{\rho \nu} & (G_E^{-1})^{\mu \nu} - \kappa^2 (\theta G_E \theta)^{\mu \nu} 
\end{pmatrix} \, .
\end{equation}

The T-duality relations (\ref{eq:xcong}) rewritten in terms of canonical variables are
\begin{subequations}\label{eq:relx}
\begin{eqnarray}\label{eq:dotx1}
\label{eq:xprime}
x^{\prime\mu}&\cong&
\frac{1}{\kappa}\,^\star\pi^\mu
-\kappa\theta_{0}^{\mu\nu}\beta^{0}_\nu[V] \, ,
\\
\pi_\mu&\cong&
\kappa y^\prime_\mu+\kappa\beta^{0}_\mu[V] \, ,
\end{eqnarray}
\end{subequations} 
and the inverse T-duality relations (\ref{eq:ycong}) by
\begin{subequations}\label{eq:relycan}
\begin{eqnarray}
\label{eq:yprimecan}
y^\prime_\mu&\cong&
\frac{1}{\kappa}\pi_\mu-\beta^{0}_\mu[x],
\\
\label{eq:starp}
^\star\pi^\mu&\cong&
\kappa x^{\prime\mu}+\kappa^{2}\theta_{0}^{\mu\nu}\beta^{0}_\nu[x].
\end{eqnarray}
\end{subequations}
The T-duality relation (\ref{eq:yprimecan}) mixes the coordinate and momenta, which is the source of non-commutativity between the T-dual coordinates. The full Poisson bracket relations between T-dual phase space were obtained in \cite{NCwcb}. They are given by\begin{subequations}
\begin{eqnarray} \label{eq:NC1}
\{ y^{\prime}_\mu (\sigma), y^{\prime}_\nu (\bar{\sigma}) \} &\cong& \frac{1}{\kappa} B_{\mu \nu \rho} x^{\prime \rho} \delta(\sigma-\bar{\sigma}) \\  \label{eq:NC2}
\{^\star\pi^\mu(\sigma),\,^\star\pi^\nu(\bar{\sigma})\}&\cong& 0 \\ \label{eq:NC3}
\{ y^{\prime}_\mu (\sigma), {}^\star \pi^\nu (\bar{\sigma}) \} &\cong &  (\delta^\nu_\mu + \kappa \theta_0^{\nu \rho} h_{\rho \mu})\delta^{\prime}(\sigma-\bar{\sigma}) +\frac{2 \kappa}{3}\theta^{\nu \rho}_0 B_{\rho \mu \sigma } x^{\prime \sigma} \delta(\sigma-\bar{\sigma}) \, ,
\end{eqnarray}
\end{subequations}
where throughout the paper we assume the notation 
\begin{equation}
\delta^{\prime }(\sigma-\bar{\sigma}) = \frac{\partial}{\partial \sigma} \delta(\sigma-\bar{\sigma}) \, .
\end{equation}
The flux terms in (\ref{eq:NC1}) arise from infinitesimal corrections of order $h_{\mu  \nu}$, which appears through $\beta$-function terms in (\ref{eq:relycan}). 

\section{Generalized current algebra and the T-duality-induced \\ 
Courant bracket}
\cleq

We want to obtain the analogue of the Courant bracket in the T-dual description. We will consider the generalized charges defined in the basis of T-dual variables ${}^\star \pi^\mu$ and $\kappa y_\mu^{\prime}$. Analogously to (\ref{eq:G-def}), we define the generalized charges by
\begin{equation} \label{eq:Gstar-def}
{}^{\star} {\cal Q}_{\hat{\Lambda}} = \int_0^{2\pi} d\sigma\  \hat{\Lambda}^M \eta_{MN}\ {}^\star X^N = \int_0^{2\pi} d\sigma \Big( \hat{\xi}^\mu (V (\sigma)) \kappa y_{\mu}^{\prime} + \hat{\lambda}_\mu (V (\sigma))\ {{}^\star \pi}^\mu \Big) \, ,
\end{equation}
where
\begin{equation} \label{eq:Lstar-def}
\hat{\Lambda}^M (V (\sigma))= \begin{pmatrix}
\hat{\xi}^\mu (V (\sigma))\\
\hat{\lambda}_\mu ( V(\sigma))
\end{pmatrix}
\, .
\end{equation}
We choose the parameters that depend on the line integral $V^\mu$. It is worth taking some time to motivate this choice. To impose that the parameters $\hat{\Lambda}$ depend solely on the T-dual coordinate seems inadequate - the fact that the T-dual description is nonlocal, indicate that we should look for the charge parameters that are also nonlocal. Moreover, the T-dual coordinates do not commute, meaning that such type of generalized charges would have non-trivial brackets between the parameters, leading to the algebra that does not close. 

In order to obtain equal time Poisson bracket relations between charges, we apply the T-dual relations
\begin{equation} \label{eq:G-star-dual}
{}^{\star} {\cal Q}_{\hat{\Lambda}} \cong \int_0^{2\pi} d\sigma \Big( \hat{\xi}^\mu (x) i_{\mu} + \hat{\lambda}_\mu (x) l^\mu \Big) \, ,
\end{equation}
where we have introduced two auxiliary currents  $\kappa y^{\prime}_\mu \cong i_\mu \, ,  \ {}^\star \pi^\mu \cong l^\mu$
\begin{eqnarray} \label{eq:aux-cur0}
i_\mu &=& \pi_\mu- \kappa h_{\mu \nu} x^{\prime \nu} \, ,
\\ \label{eq:aux-cur00}
l^\mu &=& \kappa x^{\prime\mu}+\kappa^{2}\theta_{0}^{\mu\nu} h_{\nu \rho} x^{\prime \rho} \, . 
\end{eqnarray}
They can easily be inverted, taking into account weakly curved approximation
\begin{eqnarray}  \label{eq:aux-cur1}
\kappa x^{\prime\mu} &=&(\delta^\mu_{\ \nu} - \kappa\theta_0^{\mu\rho}h_{\rho\nu}) l^\nu \, , \\  \label{eq:aux-cur11}
\pi_\mu &=& i_\mu+h_{\mu\nu}l^\nu \, .
\end{eqnarray}
Using (\ref{eq:NC1}) - (\ref{eq:NC3}) , we can easily write
\begin{subequations}
\begin{eqnarray} \label{eq:NC-c1}
\{ i_\mu (\sigma), i_\nu (\bar{\sigma}) \} &=&  B_{\mu \nu \rho} l^{\rho} \delta(\sigma-\bar{\sigma})  \, , \\  \label{eq:NC-c2}
\{l^\mu(\sigma),l^\nu(\bar{\sigma})\}&=& 0 \, ,  \\ \label{eq:NC-c3}
\{ i_\mu (\sigma), l^\nu (\bar{\sigma}) \} &=&  \kappa(\delta^\nu_\mu + \kappa \theta_0^{\nu \rho} h_{\rho \mu})\delta^{\prime}(\sigma-\bar{\sigma}) +\frac{2 \kappa}{3}\theta^{\nu \rho}_0 B_{ \rho \mu \sigma} l^{\sigma} \delta(\sigma-\bar{\sigma}) \, .
\end{eqnarray}
\end{subequations}
where we used (\ref{eq:aux-cur0}) and the fact that quadratic terms in $h$ are negligible in the weakly curved approximation.

In addition to this, we will need a few additional auxiliary Poisson bracket relations involving $V^\mu$. First, using (\ref{eq:x-cong-v}), we obtain
\begin{equation} \label{eq:auxPB0}
\{ V^\mu (\sigma), V^\nu (\bar{\sigma}) \} \cong \{ x^\mu(\sigma) , x^\nu (\bar{\sigma})\} = 0 \, ,
\end{equation}
which consequently implies 
\begin{equation} \label{eq:auxPB1}
\{ \hat{\Lambda}^M_1 (V (\sigma)), \hat{\Lambda}^N_2 (V(\bar{\sigma})) \} \cong 0 \, .
\end{equation}
Next, we obtain
\begin{equation}\label{eq:auxPB20}
\begin{aligned}
\left\{
V^\mu(\sigma),
{}^\star\pi^\nu(\bar{\sigma})
\right\}
&\cong
\left\{
x^\mu(\sigma) ,
\kappa x^{\prime\nu}(\bar{\sigma})
+
\kappa^2\theta_0^{\nu\rho}
\beta_\rho^0[x(\bar{\sigma})]
\right\}=0 \, ,
\end{aligned}
\end{equation}
which results in 
\begin{equation}\label{eq:auxPB2}
\begin{aligned}
\left\{
\hat{\Lambda}^M(V(\sigma)),
{}^\star\pi^\nu(\bar{\sigma})
\right\}
&\cong
0\, ,
\end{aligned}
\end{equation}

Lastly, we have
\begin{equation} \label{eq:auxPB30}
\begin{aligned}
\left\{
 V^\mu(\sigma),
\kappa y_\nu^\prime(\bar{\sigma})
\right\}
&\cong
\left\{
x^\mu(\sigma),
\pi_\nu(\bar{\sigma})
-
\kappa\beta_\nu^0[x(\bar{\sigma})]
\right\}
\\
&=
\delta^\mu_{\ \nu} \delta(\sigma-\bar{\sigma})\, ,
\end{aligned}
\end{equation}
or equivalently 
\begin{equation} \label{eq:auxPB3}
\begin{aligned}
\left\{
\hat{\Lambda}^M(V(\sigma)),
\kappa y_\nu^\prime(\bar{\sigma})
\right\}
&\cong
\partial_\nu \hat{\Lambda}^M \delta(\sigma-\bar{\sigma})\, .
\end{aligned}
\end{equation}

Now we have all necessary formulae to obtain the Poisson bracket algebra between two T-dual generalized charges (\ref{eq:G-star-dual}). We will calculate term by term. Firstly, using (\ref{eq:NC-c1}), (\ref{eq:auxPB1}) and (\ref{eq:auxPB3}), we obtain
\begin{eqnarray} \label{eq:aux11}
\{ \hat{\xi}_1^\mu (\sigma) i_\mu (\sigma), \hat{\xi}_2^\nu (\bar{\sigma}) i_\nu (\bar{\sigma}) \} = -(\hat{\xi}_1^\nu \partial_\nu \hat{\xi}_2^\mu - \hat{\xi}_2^\nu \partial_\nu \hat{\xi}_1^\mu) i_\mu \delta(\sigma-\bar{\sigma}) + B_{\mu \nu \rho} \hat{\xi}_1^\nu \hat{\xi}_2^\rho l^\mu \delta(\sigma-\bar{\sigma})\, . 
\end{eqnarray}
From (\ref{eq:NC-c2}), (\ref{eq:auxPB1}) and (\ref{eq:auxPB2}), we obtain 
\begin{eqnarray} \label{eq:aux12}
\{ \hat{\lambda}_{1\mu} (\sigma) l^\mu (\sigma), \hat{\lambda}_{2\nu} (\bar{\sigma}) l^\nu (\bar{\sigma}) \} =0 \, . 
\end{eqnarray}
For the mixed term, using (\ref{eq:NC-c3}), (\ref{eq:auxPB1}), (\ref{eq:auxPB2}) and (\ref{eq:auxPB3}), we write
\begin{eqnarray} \label{eq:aux131}
\{ \hat{\xi}_1^\mu (\sigma) i_\mu (\sigma), \hat{\lambda}_{2\nu} (\bar{\sigma}) l^\nu (\bar{\sigma}) \} = -(\hat{\xi}_1^\nu \partial_\nu \hat{\lambda}_{2\mu} - \frac{2\kappa}{3} \theta_0^{\nu \rho} B_{\rho \sigma \mu} \hat{\xi}_1^{\sigma} \hat{\lambda}_{2\nu}) l^\mu \delta(\sigma-\bar{\sigma}) + {\cal A}(\sigma, \bar{\sigma}) \, , 
\end{eqnarray}
where ${\cal A}$ is the anomalous part, given by
\begin{eqnarray} \label{eq:aux1an}
{\cal A}(\sigma, \bar{\sigma}) &=& \kappa \hat{\xi}_1^\mu (\sigma) \hat{\lambda}_{2\nu}(\bar{\sigma}) (\delta^\nu_\mu + \kappa \theta_0^{\nu \rho} h_{\rho \mu} (\sigma)) \delta^{\prime}(\sigma-\bar{\sigma}) \, .
\end{eqnarray}
In order to deal with the anomalous part we will need following $\delta$-function identities:
\begin{equation} \label{eq:delta1}
\delta^{\prime} (\sigma-\bar{\sigma}) = \frac{1}{2} \Big( \delta^{\prime} (\sigma-\bar{\sigma}) - \frac{\partial}{\partial \bar{\sigma}}\delta (\sigma-\bar{\sigma}) \Big) \, , 
\end{equation}
and
\begin{equation}\label{eq:delta2}
f(\bar{\sigma}) \delta^{\prime} (\sigma-\bar{\sigma}) = f(\sigma) \delta^{\prime} (\sigma-\bar{\sigma}) + f^{\prime}(\sigma) \delta (\sigma-\bar{\sigma}) \, .
\end{equation}
After subsequent application of these relations, the anomalous part (\ref{eq:aux1an}) becomes
\begin{eqnarray} \label{eq:auxAN}
{\cal A}(\sigma, \bar{\sigma}) &=& \frac{\kappa}{2}\hat{\xi}_1^\mu \hat{\lambda}_{2\nu} (\delta^\nu_\mu + \kappa \theta_0^{\nu \rho} h_{\rho \mu} ) \delta^{\prime}(\sigma-\bar{\sigma}) \\ \notag
&&- \frac{\kappa}{2}\hat{\xi}_1^\mu(\bar{\sigma}) \hat{\lambda}_{2\nu}(\bar{\sigma}) (\delta^\nu_\mu + \kappa \theta_0^{\nu \rho} h_{\rho \mu} (\bar{\sigma})) \frac{\partial}{\partial \bar{\sigma}}\delta (\sigma-\bar{\sigma}) \\ \notag
&& + \frac{\kappa}{2} x^{\prime \mu} \Big(\hat{\xi}_1^\nu \partial_\mu \hat{\lambda}_{2\nu} - \partial_\mu \hat{\xi}_1^\nu \hat{\lambda}_{2\nu} + \kappa \hat{\xi}_1^\rho (h \theta_0)_\rho^{\ \nu} \partial_\mu \hat{\lambda}_{2\nu} - \kappa \hat{\lambda}_{2\nu} \partial_\mu  (\hat{\xi}_1 h \theta_0)^{\ \nu}\Big) \delta(\sigma-\bar{\sigma}) \\ \notag
&=& {\cal A}_0 (\sigma, \bar{\sigma}) \\ \notag
&& + \kappa x^{\prime \mu} \Big(\hat{\xi}_1^\nu \partial_\mu \hat{\lambda}_{2\nu} - \frac{1}{2} \partial_\mu (\hat{\xi}_1 \hat{\lambda}_{2}) + \kappa \hat{\xi}_1^\rho (h \theta_0)_\rho^{\ \nu}\partial_\mu \hat{\lambda}_{2\nu} - \frac{\kappa}{2} \partial_\mu  (\hat{\xi}_1 h \theta_0 \hat{\lambda}_{2})\Big) \delta(\sigma-\bar{\sigma}) \, ,
\end{eqnarray}
where we used $(h\theta_0)_\mu^{\ \nu} = (\theta_0 h)_{\ \mu}^{\nu}$, and in the last step we applied the chain rule, and defined 
\begin{eqnarray} \label{eq:auxA0}
{\cal A}_0 (\sigma, \bar{\sigma}) &=& \frac{\kappa}{2}\hat{\xi}_1^\mu \hat{\lambda}_{2\nu} (\delta^\nu_\mu + \kappa \theta_0^{\nu \rho} h_{\rho \mu} ) \delta^{\prime}(\sigma-\bar{\sigma})  \\ \notag
&&- \frac{\kappa}{2}\hat{\xi}_1^\mu(\bar{\sigma}) \hat{\lambda}_{2\nu}(\bar{\sigma}) (\delta^\nu_\mu + \kappa \theta_0^{\nu \rho} h_{\rho \mu} (\bar{\sigma})) \frac{\partial}{\partial \bar{\sigma}}\delta (\sigma-\bar{\sigma})  \, .
\end{eqnarray}
This anomaly will disappear after the integration over both $\sigma$ and $\bar{\sigma}$ over closed world-sheet. We would like to express the result in terms of basis currents, so substituting (\ref{eq:aux-cur1}) into (\ref{eq:auxAN}) and disregarding all terms of quadratic order of $h$, we obtain 
\begin{eqnarray} \label{eq:aux1an1}
{\cal A}(\sigma, \bar{\sigma}) &=& {\cal A}_0 (\sigma, \bar{\sigma})  \\ \notag
&& +\Big(\hat{\xi}_1^\nu \partial_\mu \hat{\lambda}_{2\nu} - \frac{1}{2} \partial_\mu (\hat{\xi}_1 \hat{\lambda}_2) - \kappa (h\theta_0)_\mu^{\ \rho}(\hat{\xi}_1^\nu \partial_\rho \hat{\lambda}_{2\nu}- \frac{1}{2} \partial_\rho (\hat{\xi}_1 \hat{\lambda}_2)) \\ \notag
&&+ \kappa (h \theta_0)_\rho^{\ \nu} (\hat{\xi}_1^\rho \partial_\mu \hat{\lambda}_{2\nu} - \frac{1}{2}\partial_\mu(\hat{\xi}_1^\rho \hat{\lambda}_{2\nu})) - \frac{\kappa}{6} \theta_0^{\nu \rho } B_{\rho \sigma \mu }\hat{\xi}_1^\sigma  \hat{\lambda}_{2\nu} \Big) l^\mu \delta(\sigma-\bar{\sigma}) \, .
\end{eqnarray}
The second mixed term can be calculated from antisymmetric properties of Poisson bracket
\begin{equation}
\{ \hat{\lambda}_{1\mu} (\sigma) l^\mu (\sigma), \hat{\xi}_2^\nu (\bar{\sigma}) i_\nu (\bar{\sigma})\} = - \{\hat{\xi}_2^\nu (\bar{\sigma}) i_\nu (\bar{\sigma}), \hat{\lambda}_{1\mu} (\sigma) l^\mu (\sigma)\} \, .
\end{equation}

After some straightforward calculations, we obtain 
\begin{equation} \label{eq:Gstar-aux}
\{ {}^{\star} {\cal Q}_{\hat{\Lambda}_1} , {}^{\star} {\cal Q}_{\hat{\Lambda}_2}  \} =- {}^{\star} {\cal Q}_{[\hat{\Lambda}_1, \hat{\Lambda}_2]_{\hat{C}}} \, , 
\end{equation}
where the algebra closes on weakly curved generalisation of the Courant bracket. If we assume notation ${}^{\star} {\cal Q}_{[\hat{\Lambda}_1, \hat{\Lambda}_2]_{\hat{C}}} \cong \int d\sigma \Big( \hat{\xi}^\mu (x) i_\mu + \hat{\lambda}_\mu(x) l^\mu \Big) $, the resulting parameters are given by 
\begin{eqnarray} \label{eq:Courant-dual10}
\hat{\xi}^\mu = \hat{\xi}^\nu_1 \partial_\nu \hat{\xi}_2^\mu - \hat{\xi}^\nu_2 \partial_\nu \hat{\xi}_1^\mu \, ,
\end{eqnarray}
and
\begin{align} \label{eq:Courant-dual11}
\hat{\lambda}_{\mu}
={}&
\hat{\xi}_1^\nu
\left(
\partial_\nu\hat{\lambda}_{2\mu}
-\partial_\mu\hat{\lambda}_{2\nu}
\right)
-\hat{\xi}_2^\nu
\left(
\partial_\nu\hat{\lambda}_{1\mu}
-\partial_\mu\hat{\lambda}_{1\nu}
\right)+ \frac{1}{2} \partial_\mu (\hat{\xi}_1^\nu  \hat{\lambda}_{2\nu} - \hat{\xi}_2^\nu \hat{\lambda}_{1\nu})
\nonumber\\
&-\kappa\theta_0^{\rho\sigma}h_{\sigma \nu}
(
\hat{\xi}_1^\nu\partial_\mu\hat{\lambda}_{2\rho}
-\hat{\xi}_2^\nu\partial_\mu\hat{\lambda}_{1\rho}
 - \frac{1}{2} \partial_\mu (\hat{\xi}_1^\nu \hat{\lambda}_{2\rho}-\hat{\xi}_2^\nu \hat{\lambda}_{1\rho}) )
\nonumber\\
&+\kappa\theta_0^{\rho\sigma }h_{\sigma \mu}(
\hat{\xi}_1^\nu\partial_\rho\hat{\lambda}_{2\nu}
-\hat{\xi}_2^\nu\partial_\rho\hat{\lambda}_{1\nu}
 - \frac{1}{2} \partial_\rho (\hat{\xi}_1^\tau \hat{\lambda}_{2\tau}-\hat{\xi}_2^\tau \hat{\lambda}_{1\tau}) )
\nonumber\\
&-B_{\mu\nu\rho}\,
\hat{\xi}_1^\nu\hat{\xi}_2^\rho
-\frac{\kappa}{2}
\theta_0^{\nu \rho}B_{\rho \sigma \mu}
\left(
\hat{\xi}_1^\sigma \hat{\lambda}_{2\nu}
-\hat{\xi}_2^\sigma \hat{\lambda}_{1\nu}
\right) .
\end{align}

To make the geometric structure of the resulting bracket explicit, we now rewrite it in coordinate-free notation. The vector component is the Lie bracket, while the one-form component can be decomposed as
\begin{equation}
\hat{\lambda} = \hat{\lambda}^{c} + \hat{\lambda}^{h}  - dB(\hat{\xi}_1, \hat{\xi}_2, .)  \, , 
\end{equation}
where $\hat{\lambda}^c$ is one-form obtained in the standard Courant bracket, $ dB(\hat{\xi}_1, \hat{\xi}_2, .)  $ represent the pure flux term contracted with components of two vector fields, and $\hat{\lambda}^{h}$ is auxiliary one-form that explicitly depend on $h$ that we are yet to determine. We obtain
\begin{eqnarray} \label{eq:lambda-h}
\hat{\lambda}^{h}_\mu &=&-\kappa h_{\sigma \nu}(\hat{\xi}_1^\nu\partial_\mu(\hat{\lambda}_{2}\theta_0)^\sigma-\frac{1}{2}\partial_\mu(\hat{\xi}_1^\nu (\hat{\lambda}_{2}\theta_0 )^\sigma)) -\frac{\kappa}{2} B_{\rho\nu\mu}\hat{\xi}_1^\nu(\hat{\lambda}_{2}\theta_0)^\rho \\ \notag
&&+\kappa\theta_0^{\rho\sigma }h_{\sigma \mu}(
\hat{\xi}_1^\nu\partial_\rho\hat{\lambda}_{2\nu}- \frac{1}{2} \partial_\rho (\hat{\xi}_1 \hat{\lambda}_2))- (1 \leftrightarrow 2) \\ \notag
&=& -\kappa h_{\sigma \nu}(\hat{\xi}_1^\nu\partial_\mu(\hat{\lambda}_{2}\theta_0)^\sigma-\frac{1}{2}\partial_\mu(\hat{\xi}_1^\nu (\hat{\lambda}_{2}\theta_0 )^\sigma)) -\frac{\kappa}{2} B_{\rho\nu\mu}\hat{\xi}_1^\nu(\hat{\lambda}_{2}\theta_0)^\rho \\ \notag
&&+\kappa\theta_0^{\rho\sigma }h_{\sigma \mu}(
\hat{\xi}_1^\nu(\partial_\rho\hat{\lambda}_{2\nu}-\partial_\nu \hat{\lambda}_{2\rho})- \frac{1}{2} \partial_\rho (\hat{\xi}_1 \hat{\lambda}_2))+\kappa h_{\sigma \mu} \hat{\xi}_1^\nu \partial_\nu (\hat{\lambda}_{2}\theta_0)^\sigma- (1 \leftrightarrow 2)\\ \notag
&=&\kappa \Big[ \hat{\xi}_1^\nu \Big(\partial_\nu (\hat{\lambda}_2 \theta_0 h)_\mu - \partial_\mu (\hat{\lambda}_2 \theta_0 h)_\nu \Big) + \frac{1}{2} \partial_\mu (\hat{\lambda}_2 \theta_0 h \hat{\xi}_1) \\ \notag
&&- \hat{\xi}_1^\nu (\hat{\lambda}_2 \theta_0)^\sigma (\partial_\nu h_{\sigma \mu} - \partial_\mu h_{\sigma \nu} + \frac{1}{2}\partial_\mu h_{\sigma \nu} + \frac{1}{2}B_{\sigma \nu \mu})- (1 \leftrightarrow 2) \Big] \\ \notag
&&-\kappa\hat{\lambda}^c_\rho \theta_0^{\rho \sigma} h_{\sigma \mu} \\ \notag
&=& \Big[ \kappa \hat{\xi}_1^\nu \Big(\partial_\nu (\hat{\lambda}_2 \theta_0 h)_\mu - \partial_\mu (\hat{\lambda}_2 \theta_0 h)_\nu\Big) + \frac{\kappa}{2} \partial_\mu (\hat{\lambda}_2 \theta_0 h \hat{\xi}_1)- (1 \leftrightarrow 2) \Big]-\kappa\hat{\lambda}^c_\rho \theta_0^{\rho \sigma} h_{\sigma \mu} \, ,
\end{eqnarray} 
where in the first line we just used Leibniz rule and contracted some indices. In the second line, we added and subtracted the term $\kappa h_{\sigma \mu} \hat{\xi}_1^\nu \partial_\nu (\hat{\lambda}_{2}\theta_0)^\sigma - (1 \leftrightarrow 2)$. In the third line we recognized the term containing $\hat{\lambda}_c$ and applied Leibniz rule once again. Lastly, we used the properties of the Kalb-Ramond field, which results in canceling of the terms containing its field strength. Now we can write the weakly curved generalization of the Courant bracket in non-coordinate notation
\begin{eqnarray} \label{eq:gen-courant-nc}
\hat{\xi} &=& [\hat{\xi}_1, \hat{\xi}_2]_L \\ \notag
\hat{\lambda} &=& {\cal{L}}_{\hat{\xi}_1} \hat{\lambda}_2 - {\cal{L}}_{\hat{\xi}_2} \hat{\lambda}_1 - \frac{1}{2}d (i_{\hat{\xi}_1} \hat{\lambda}_2 - i_{\hat{\xi}_2} \hat{\lambda}_1) \\ \notag
&& +\kappa \Big({\cal L}_{\hat{\xi}_1}(i_{\hat{\lambda}_2 \theta_0}h)-{\cal L}_{\hat{\xi}_2}(i_{\hat{\lambda}_1 \theta_0}h) - \frac{1}{2}d(i_{\hat{\xi}_1} i_{\hat{\lambda}_2 \theta_0} h -i_{\hat{\xi}_2} i_{\hat{\lambda}_1 \theta_0} h) - i_{\hat{\lambda}^c \theta_0} h \Big) - dB(\hat{\xi}_1, \hat{\xi}_2, .) \, .
\end{eqnarray}

\subsection{Courant algebroid interpretation and isomorphism}

We can rewrite these relations in the form more appropriate for generalized geometry. The currents can be written as the generalized current vector $\hat{X}^M = (l^\mu , i_\mu)^T$, which can be expressed by
\begin{equation} \label{eq:hatX-def}
\hat{X}^M = T^M_{\ N} X^N \, , \quad T^M_{\ N}  = \begin{pmatrix}
\delta^\mu_\nu + \kappa \theta_0^{\mu \rho} h_{\rho \nu}& 0 \\
- h_{\mu \nu} & \delta_\mu^\nu 
\end{pmatrix} \, ,
\end{equation}
where $X^M$ is defined in (\ref{eq:XM-def}). The charge T-duality relation (\ref{eq:G-star-dual}) can be rewritten as
\begin{equation} \label{eq:G-star-dual1}
^\star {\cal Q}_{\hat{\Lambda}} \cong \int d\sigma \hat{\Lambda}^M \eta_{MN} \hat{X}^N = \int d\sigma \hat{\Lambda}^M \eta_{MN} T^N_{\ P} X^P \, .
\end{equation}
Let us also define another transformation by
\begin{equation} \label{eq:F-def}
F =  \eta T^T \eta  \, , \quad  F^M_{\ N}=
\begin{pmatrix}
\delta^\mu{}_\nu
&
0
\\[1mm]
h_{\mu\nu}
&
\delta_\mu{}^\nu
+
\kappa h_{\mu\rho}\theta_0^{\rho\nu}
\end{pmatrix}\, ,
\end{equation}
so that we can define another set of parameters by
\begin{equation} \label{eq:LFLhat}
\Lambda^M = F^M_{\ N} \hat{\Lambda}^N \, .
\end{equation}
Its inverse is given by
\begin{equation} \label{eq:Finv-def}
(F^{-1})^M_{\ N}=
\begin{pmatrix}
\delta^\mu{}_\nu
&
0
\\[1mm]
-h_{\mu\nu}
&
\delta_\mu{}^\nu
-
\kappa h_{\mu\rho}\theta_0^{\rho\nu}
\end{pmatrix}\, ,
\end{equation}

This transformation is, however, not $O(D, D)$ invariant for the standard inner product, since $F^T \eta F \neq  \eta$.
The relation (\ref{eq:G-star-dual1}) now becomes
\begin{equation} \label{eq:G-star-dual2}
^\star {\cal Q}_{\hat{\Lambda}} \cong \int d\sigma \Lambda^M \eta_{MN} X^N = {\cal Q}_\Lambda \, ,
\end{equation}
whose charge algebra is known to produce Courant bracket. Using (\ref{eq:G-star-dual2}) and comparing (\ref{eq:Gstar-aux}) and (\ref{eq:GG-courant}), we obtain
\begin{equation} \label{eq:LL-F-aux}
([\hat{\Lambda}_1, \hat{\Lambda}_2]_{\hat{C}})^M \eta_{MN} \hat{X}^N = ([\Lambda_1, \Lambda_2]_{\cal C})^M \eta_{MN} X^N \, .
\end{equation}
This relation holds for the Courant representatives fixed by this common symmetrization prescription. The left hand side can be transported using (\ref{eq:G-star-dual2}) into
\begin{equation}
( F [\hat{\Lambda}_1, \hat{\Lambda}_2]_{\hat{C}})^M \eta_{MN} X^N \, ,
\end{equation}
which after substituting (\ref{eq:LFLhat}) into the right hand side of (\ref{eq:LL-F-aux}) gives the following relation
\begin{equation} \label{eq:hatC-C}
[\hat{\Lambda}_1, \hat{\Lambda}_2]_{\hat{C}} = F^{-1}[F \hat{\Lambda}_1, F \hat{\Lambda}_2]_{\cal C} \, . 
\end{equation}
We see that the weakly curved generalization of the Courant bracket is isomorphic to the standard Courant bracket with transformation F (\ref{eq:F-def}) acting as isomorphism. In fact, the whole Courant algebroid structure can be transported by this isomorphism (\ref{eq:CA-isom}). The appropriate inner product is pullback of the original inner product
\begin{equation} \label{eq:eta-F}
\eta_F = F^T \eta F \, , \quad (\eta_F)_{MN} = \begin{pmatrix}
0 & \delta_\mu^\nu+ \kappa h_{\mu \rho} \theta_0^{\rho \nu} \\ 
\delta_\nu^\mu + \kappa \theta_0^{ \mu \rho} h_{\rho \nu}  & 0 
\end{pmatrix} \, .
\end{equation}
The anchor is obtained from relation $\rho_F= \rho \circ F$, which remains the projection on tangent space $\rho_0$ in this specific choice of $F$. Therefore, we obtain the following isomorphism between the two Courant algebroid structures
\begin{equation}
F: (TM \oplus T^{\star}M , [,]_{\hat{C}}, \eta_F, \rho_0) \to (TM \oplus T^{\star}M , [,]_{{\cal C}}, \eta, \rho_0) \, .
\end{equation}

Finally, we interpret the relation between the initial and T-dual generalized-charge algebras, extending the current algebra viewpoint of \cite{cdual}. The definition of the T-dual charge \({}^\star{\cal Q}_{\hat\Lambda}\) (\ref{eq:Gstar-def}) is obtained from that of the initial charge \({\cal Q}_{\Lambda}\) (\ref{eq:G-def}) by formally replacing the variables of the initial theory with their T-dual counterparts
\begin{equation}
\kappa x^{\prime\mu}\leftrightarrow\kappa y'_\mu \, ,
\qquad
\pi_\mu\leftrightarrow{}^\star\pi^\mu \, ,
\end{equation}
in the current bases. These replacements are accompanied by the exchange of vector and one-form parameters and the replacement of their argument $x$ by $V$. Their Poisson bracket algebras are governed by the standard Courant bracket and its weakly curved generalization, respectively. The map \(F\) established above relates these brackets and, together with the transported pairing and anchor, defines an isomorphism between the associated Courant algebroid structures. In this sense, the weakly curved generalization provides the T-dual realization of the standard Courant bracket.

\section{The $B$-frame of the T-dual theory}
\cleq

We will define the noncanonical currents in $B$-frame by 
\begin{equation}
{}^\star J_{\hat{B}}^M = ( e^{\hat{B}})_{\ N}^{M}\ {}^\star X^N = \begin{pmatrix}
{}^\star \pi^\mu  \\
\kappa y_{\mu}^{\prime} + 2 B_{\mu \nu}[V] {}^\star \pi^\nu \, ,
\end{pmatrix} \equiv \begin{pmatrix}
{}^\star \pi^\mu \\
{}^\star j_{\mu}
\end{pmatrix} \, .
\end{equation}
Let us define 
${}^\star J_{\hat{B}}^M \cong J_{\hat{B}}^M \, , {}^\star \pi^\mu  \cong l^\mu \, ,  \ {}^\star j_{\mu} \cong j_\mu $, where $l^\mu$ is defined in (\ref{eq:aux-cur00}), while $j_\mu$ is as easily obtained from (\ref{eq:yprimecan})  
\begin{eqnarray} \label{eq:j-def}
j_{\mu} =  \pi_\mu + \kappa{\cal B}_{\mu \nu}x^{\prime \nu}, \quad {\cal B}_{\mu \nu} = 2 b_{\mu \nu} +  h_{\mu \nu} +2\kappa (b\theta_0)_\mu^{\ \rho} h_{\rho \nu} \, .
\end{eqnarray}
The equal-time Poisson bracket algebra of currents is 
\begin{equation}
\begin{aligned}
\left\{
j_\mu(\sigma),
j_\nu(\bar{\sigma})
\right\}
={}&
\kappa
\left[
{\cal B}_{\mu\nu}(\sigma)
+
{\cal B}_{\nu\mu}(\bar{\sigma})
\right]
\delta^{\prime}(\sigma-\bar{\sigma})
\\
&+
\kappa
\left(
\partial_\nu {\cal B}_{\mu\rho}
-
\partial_\mu {\cal B}_{\nu\rho}
\right)
x^{\prime\rho}(\sigma)
\delta(\sigma-\bar{\sigma}) \, .
\end{aligned}
\end{equation}
In calculating algebra of currents one has to be careful - since $b_{\mu \nu}$ and $h_{\mu \nu}$ are both antisymmetric, but $(b\theta_0  h)_{\mu \nu}$ is not.  After applying the $\delta$-function identity (\ref{eq:delta2}), and $\partial_\mu h_{\nu \rho} =\frac{1}{3} B_{\mu \nu \rho}$, and using $l^\mu = \kappa x^{\prime \mu} + {\cal O}(h)$ (\ref{eq:aux-cur00}), one obtains the algebra of currents 
\begin{eqnarray}\label{eq:NC-B1}
\{j_{\mu}(\sigma),j_{\nu}(\bar{\sigma})\}
&=&
2\kappa^2
\bigl(b\theta_0h-h\theta_0b\bigr)_{\mu\nu}
\delta'(\sigma-\bar{\sigma})
\nonumber\\
&&+
\Big[
-B_{\mu\nu\rho}
-\frac{2\kappa}{3}
(b\theta_0)_{\mu}{}^{\sigma }B_{\sigma \nu\rho}
+\frac{4\kappa}{3}
(b\theta_0)_{\nu}{}^{\sigma }B_{\sigma \mu\rho}
\Big]
l^\rho \,
\delta(\sigma-\bar{\sigma}) \, .
\end{eqnarray}
In a similar way, one obtains the relation
\begin{eqnarray} \label{eq:NC-B2}
\{ j_\mu (\sigma), l^\nu (\bar{\sigma}) \} &=&  \kappa(\delta^\nu_\mu + \kappa \theta_0^{\nu \rho} h_{\rho \mu})\delta^{\prime}(\sigma-\bar{\sigma}) +\frac{2 \kappa}{3}\theta^{\nu \rho}_0 B_{ \rho \mu \sigma} l^{\sigma} \delta(\sigma-\bar{\sigma}) \, ,
\end{eqnarray}
which together with trivial relation (\ref{eq:NC-c2}) constitute all relations between noncanonical currents in $B$-frame.

\subsection{Generalized charge algebra and the $B$-twisted bracket}

We define
\begin{eqnarray} \label{eq:GB-star-dual}
&& {}^{\star} {\cal Q}^{B}_{\hat{\Lambda}} = \int_0^{2\pi} d\sigma\  \hat{\Lambda}^M \eta_{MN}\ {}^\star J_{\hat{B}}^N \, \\ \notag
&& {}^{\star} {\cal Q}^{B}_{\hat{\Lambda}} \cong \int_0^{2\pi} d\sigma \Big( \hat{\xi}^\mu (x) j_{\mu} + \hat{\lambda}_\mu (x) l^\mu \Big) \, ,
\end{eqnarray}
where $\hat{\Lambda}^M$ has the same form as in (\ref{eq:Lstar-def}). In addition to the relations (\ref{eq:auxPB1}) - (\ref{eq:auxPB3}), we will require
\begin{equation}\label{eq:auxPB4}
\begin{aligned}
\left\{
V^\mu(\sigma),
{}^\star j_\nu(\bar{\sigma})
\right\}
&\cong
\left\{
x^\mu(\sigma) ,
\pi_\nu + \kappa {\cal B}_{\nu \rho}x^{\prime \rho}\right\}=\delta^\mu_\nu \delta(\sigma-\bar{\sigma})  \, .
\end{aligned}
\end{equation}
The only contribution that differs from the previous case is
\begin{eqnarray} \label{eq:aux21}
\{ \hat{\xi}_1^\mu (\sigma) j_\mu (\sigma), \hat{\xi}_2^\nu (\bar{\sigma}) j_\nu (\bar{\sigma}) \} &=& -(\hat{\xi}_1^\nu \partial_\nu \hat{\xi}_2^\mu -  \hat{\xi}_2^\nu \partial_\nu \hat{\xi}_1^\mu) j_\mu \delta(\sigma-\bar{\sigma})  \\ \notag
&&- ( B_{\mu \nu \rho} + \frac{2\kappa}{3} (b\theta_0)_\nu^{\ \sigma} B_{\sigma \rho \mu} - \frac{4\kappa}{3} (b\theta_0)_\rho^{\ \sigma} B_{\sigma \nu \mu} )\hat{\xi}_1^\nu \hat{\xi}_2^\rho  l^\mu \delta(\sigma-\bar{\sigma})\\ \notag
&& + \tilde{\cal{A}}(\sigma, \bar{\sigma}) \, ,
\end{eqnarray}
where we have marked the anomalous part by
\begin{equation} \label{eq:aux2an}
\tilde{\cal{A}}(\sigma, \bar{\sigma}) = 2\kappa^2 (b\theta_0h(\sigma) - h(\sigma)\theta_0 b)_{\mu \nu} \hat{\xi}_1^\mu (\sigma) \hat{\xi}_2^\nu (\bar{\sigma}) \delta^{\prime}(\sigma-\bar{\sigma}) \, .
\end{equation}
In a similar manner as in the previous chapter, we write
\begin{eqnarray} \label{eq:aux2an1}
\tilde{\cal{A}}(\sigma, \bar{\sigma}) &=&  \kappa^2 (b\theta_0h(\sigma) - h(\sigma)\theta_0 b)_{\mu \nu} \hat{\xi}_1^\mu (\sigma) \hat{\xi}_2^\nu (\bar{\sigma}) (\delta^{\prime}(\sigma-\bar{\sigma})-\frac{\partial}{\partial \bar{\sigma}}\delta (\sigma-\bar{\sigma}) ) \\ \notag
&=& \tilde{\cal{A}}_0(\sigma, \bar{\sigma}) + \kappa^2 (b\theta_0 h - h\theta_0 b)_{\mu \nu}(\hat{\xi}_1^\mu \partial_\rho \hat{\xi}_2^\nu-\hat{\xi}_2^\mu \partial_\rho \hat{\xi}_1^\nu)x^{\prime \rho}\delta(\sigma-\bar{\sigma}) \\ \notag
&& - \frac{\kappa^2}{3}( (b\theta_0)_\mu^{\ \sigma} B_{\sigma \nu \rho} - B_{\mu \sigma \rho} (\theta_0 b)_{\ \nu}^{\sigma} ) \hat{\xi}_1^\mu \hat{\xi}_2^\nu x^{\prime \rho} \delta(\sigma-\bar{\sigma})\\ \notag
&=&  \tilde{\cal{A}}_0(\sigma, \bar{\sigma}) + \kappa (b\theta_0 h - h\theta_0 b)_{\mu \nu}(\hat{\xi}_1^\mu \partial_\rho \hat{\xi}_2^\nu-\hat{\xi}_2^\mu \partial_\rho \hat{\xi}_1^\nu)l^{\rho}\delta(\sigma-\bar{\sigma}) \\ \notag
&& - \frac{\kappa}{3}( (b\theta_0)_\mu^{\ \sigma} B_{\sigma \nu \rho} + (b\theta_0 )_{\nu}^{\ \sigma} B_{\sigma \mu \rho} ) \hat{\xi}_1^\mu \hat{\xi}_2^\nu l^{\rho} \delta(\sigma-\bar{\sigma}) \, , 
\end{eqnarray}
where in the first step we used (\ref{eq:delta1}), in the second step we used (\ref{eq:delta2}) and defined
\begin{eqnarray}
\tilde{\cal{A}}_0(\sigma, \bar{\sigma}) &=& \kappa^2 (b\theta_0 h - h\theta_0 b)_{\mu \nu} \hat{\xi}_1^\mu \hat{\xi}_2^\nu \delta^{\prime}(\sigma-\bar{\sigma}) \\ \notag
&&-\kappa^2 (b\theta_0 h(\bar{\sigma}) - h(\bar{\sigma})\theta_0 b)_{\mu \nu} \hat{\xi}_1^\mu (\bar{\sigma})\hat{\xi}_2^\nu(\bar{\sigma})\frac{\partial}{\partial \bar{\sigma}}\delta (\sigma-\bar{\sigma}) \, ,
\end{eqnarray}
which disappears after integration over both $\sigma$ and $\bar{\sigma}$. In the last step, we used relation (\ref{eq:aux-cur1}), together with some basic antisymmetric properties of $B_{\mu \nu \rho}$. Substituting (\ref{eq:aux2an1}) into (\ref{eq:aux21}), we obtain 
\begin{eqnarray}
\{ \hat{\xi}_1^\mu (\sigma) j_\mu (\sigma), \hat{\xi}_2^\nu (\bar{\sigma}) j_\nu (\bar{\sigma}) \} &=& \tilde{\cal{A}}_0(\sigma, \bar{\sigma}) -(\hat{\xi}_1^\nu \partial_\nu \hat{\xi}_2^\mu -  \hat{\xi}_2^\nu \partial_\nu \hat{\xi}_1^\mu) j_\mu \delta(\sigma-\bar{\sigma})\\ \notag
&& - \Big( B_{\mu \nu \rho} \hat{\xi}_1^\nu \hat{\xi}_2^\rho  - \kappa (b\theta_0 h - h\theta_0 b)_{\nu \rho} (\hat{\xi}_1^\nu \partial_\mu \hat{\xi}_2^\rho - \hat{\xi}_2^\nu \partial_\mu \hat{\xi}_1^\rho) \\ \notag
&& + \kappa ( (b\theta_0)_\nu^{\ \sigma} B_{\sigma \rho \mu} - (b\theta_0)_\rho^{\ \sigma} B_{\sigma \nu \mu})  \hat{\xi}_1^\nu \hat{\xi}_2^\rho \Big) l^\mu \delta(\sigma-\bar{\sigma}) \, .
\end{eqnarray}

Combining previous equations, we obtain
\begin{equation} \label{eq:Gstar-aux1}
\{ {}^{\star} {\cal Q}^{B}_{\hat{\Lambda}_1} , {}^{\star} {\cal Q}^{B}_{\hat{\Lambda}_2}  \} =- {}^{\star} {\cal Q}^{B}_{[\hat{\Lambda}_1, \hat{\Lambda}_2]_{\hat{C}_{B}}} \, , 
\end{equation}
where the algebra closes on B-twisted weakly curved generalisation of the Courant bracket. If we assume notation $ {}^{\star}{\cal Q}^{B}_{[\hat{\Lambda}_1, \hat{\Lambda}_2]_{\hat{C}_B}} \cong \int d\sigma \Big( \hat{\xi}^\mu (x) j_\mu + \hat{\lambda}_\mu(x) l^\mu \Big) $, the resulting parameters are given by 
\begin{eqnarray} \label{eq:Courant-dual20}
\hat{\xi}^\mu = \hat{\xi}^\nu_1 \partial_\nu \hat{\xi}_2^\mu - \hat{\xi}^\nu_2 \partial_\nu \hat{\xi}_1^\mu \, ,
\end{eqnarray}
and
\begin{align} \label{eq:Courant-dual21}
\hat{\lambda}_{\mu}
={}&
\hat{\xi}_1^\nu
\left(
\partial_\nu\hat{\lambda}_{2\mu}
-\partial_\mu\hat{\lambda}_{2\nu}
\right)
-\hat{\xi}_2^\nu
\left(
\partial_\nu\hat{\lambda}_{1\mu}
-\partial_\mu\hat{\lambda}_{1\nu}
\right)+ \frac{1}{2} \partial_\mu (\hat{\xi}_1^\nu  \hat{\lambda}_{2\nu} - \hat{\xi}_2^\nu \hat{\lambda}_{1\nu})
\nonumber\\
&-\kappa\theta_0^{\rho\sigma}h_{\sigma \nu}
(
\hat{\xi}_1^\nu\partial_\mu\hat{\lambda}_{2\rho}
-\hat{\xi}_2^\nu\partial_\mu\hat{\lambda}_{1\rho}
 - \frac{1}{2} \partial_\mu (\hat{\xi}_1^\nu \hat{\lambda}_{2\rho}-\hat{\xi}_2^\nu \hat{\lambda}_{1\rho}) )
\nonumber\\
&+\kappa\theta_0^{\rho\sigma }h_{\sigma \mu}(
\hat{\xi}_1^\nu\partial_\rho\hat{\lambda}_{2\nu}
-\hat{\xi}_2^\nu\partial_\rho\hat{\lambda}_{1\nu}
 - \frac{1}{2} \partial_\rho (\hat{\xi}_1^\nu \hat{\lambda}_{2\nu}-\hat{\xi}_2^\nu \hat{\lambda}_{1\nu}) )
\nonumber\\
&+B_{\mu\nu\rho}\,
\hat{\xi}_1^\nu\hat{\xi}_2^\rho
-\frac{\kappa}{2}
\theta_0^{\rho\alpha}B_{\alpha\nu\mu}
\left(
\hat{\xi}_1^\nu\hat{\lambda}_{2\rho}
-\hat{\xi}_2^\nu\hat{\lambda}_{1\rho}
\right) \nonumber \\
& -\kappa (b\theta_0 h - h\theta_0 b)_{\nu \rho} \left(\hat{\xi}_1^\nu \partial_\mu \hat{\xi}_2^\rho - \hat{\xi}_2^\nu \partial_\mu \hat{\xi}_1^\rho \right)  \nonumber \\
& +\kappa ( (b\theta_0)_\nu^{\ \sigma} B_{\sigma \rho \mu} - (b\theta_0)_\rho^{\ \sigma} B_{\sigma \nu \mu})  \hat{\xi}_1^\nu \hat{\xi}_2^\rho \, .
\end{align}
The difference between the non-twisted Courant bracket for T-dual theory is in the last two lines of (\ref{eq:Courant-dual21}). To write this in the coordinate-free notation, we first mark these additional terms by $\hat{\lambda}_\mu^{b}$ and write 
\begin{eqnarray} \label{eq:lambda-b1}
\hat{\lambda}_\mu^{b} &=& -\kappa (b\theta_0 h - h\theta_0 b)_{\nu \rho} \hat{\xi}_1^\nu \partial_\mu \hat{\xi}_2^\rho +\kappa (b\theta_0)_\nu^{\ \sigma} B_{\sigma \rho \mu}\hat{\xi}_1^\nu \hat{\xi}_2^\rho - (1 \leftrightarrow 2) \\ \notag
&=& -\kappa (\theta_0h)^\sigma_{\ \rho}((\hat{\xi}_1 b)_\sigma \partial_\mu \hat{\xi}_2^\rho + \partial_\mu (\hat{\xi}_2 b)_{\sigma} \hat{\xi}_1^\rho)+ \kappa \theta_0^{\nu \sigma}B_{\sigma \rho \mu} (\hat{\xi}_1 b)_\nu \hat{\xi}_2^\rho - (1 \leftrightarrow 2) \\ \notag
&=&-2\kappa (\theta_0h)^\sigma_{\ \rho} \Big(\hat{\xi}_1^\rho \partial_\mu(\hat{\xi}_2 b)_\sigma-\hat{\xi}_2^\rho  \partial_\mu(\hat{\xi}_1 b)_\sigma - \frac{1}{2}\partial_\mu (\hat{\xi}_1^\rho  (\hat{\xi}_2b)_\sigma-\hat{\xi}_2^\rho  (\hat{\xi}_1 b)_\sigma) \Big)\\ \notag
&& +\kappa \theta_0^{\nu \sigma}B_{\sigma \rho \mu} ((\hat{\xi}_1 b)_\nu \hat{\xi}_2^\rho - (\hat{\xi}_2 b)_\nu \hat{\xi}_1^\rho) \, ,
\end{eqnarray}
where in the first step we contracted appropriate vector fields with constant $b_{\mu \nu}$, followed by applying the chain rule in the next step. We further note that
\begin{eqnarray} \label{eq:lambda-b2}
&&\kappa\theta_0^{\rho\sigma }h_{\sigma \mu}(
\hat{\xi}_1^\nu\partial_\rho(\hat{\xi}_{2} b)_{\nu}
-\hat{\xi}_2^\nu\partial_\rho(\hat{\xi}_1 b)_{\nu}
 - \frac{1}{2} \partial_\rho (\hat{\xi}_1^\nu (\hat{\xi}_2 b)_{\nu}-\hat{\xi}_2^\nu (\hat{\xi}_1 b)_{\nu}) ) \\ \notag
&=& \frac{\kappa}{2}\theta_0^{\rho \sigma} h_{\sigma \mu}b_{\tau \nu} (\hat{\xi}_1^\nu \partial_\rho \hat{\xi}_2^\tau - \partial_\rho \hat{\xi}_1^\nu \hat{\xi}_2^\tau-\hat{\xi}_2^\nu\partial_\rho \hat{\xi}_1^\tau +\partial_\rho\hat{\xi}_2^\nu \hat{\xi}_1^\tau) = 0 \, ,
\end{eqnarray}
since the expression is contraction of antisymmetric tensor $b_{\tau \nu}$ and the expression depending on $\hat{\xi}$ that is symmetric in terms of $\tau, \nu$. Using  (\ref{eq:lambda-b1}) and (\ref{eq:lambda-b2}) and (\ref{eq:lambda-h}), we see that under the transformation $\hat{\lambda}_{i \mu} \to  2 \hat{\xi}_i^\nu b_{\nu \mu}$, the expression $\hat{\lambda}^h$ becomes $\hat{\lambda}_b$. Therefore, if we introduce auxiliary one-forms by
\begin{eqnarray}
\hat{\chi}_{i \mu} &=& \hat{\lambda}_{i \mu} +2 \hat{\xi}_i^\nu b_{\nu \mu} \, , \quad i = 1, 2 \\ \notag
\hat{\chi}^c &=&  {\cal{L}}_{\hat{\xi}_1} \hat{\chi}_2 - {\cal{L}}_{\hat{\xi}_2} \hat{\chi}_1 - \frac{1}{2}d (i_{\hat{\xi}_1} \hat{\chi}_2 - i_{\hat{\xi}_2} \hat{\chi}_1)	\, ,
\end{eqnarray}
the full non-coordinate expression for weakly curved generalization of the $B$-twisted Courant bracket becomes
\begin{eqnarray} \label{eq:genB-courant-nc}
\hat{\xi} &=& [\hat{\xi}_1, \hat{\xi}_2]_L \\ \notag
\hat{\lambda} &=& {\cal{L}}_{\hat{\xi}_1} \hat{\lambda}_2 - {\cal{L}}_{\hat{\xi}_2} \hat{\lambda}_1 - \frac{1}{2}d (i_{\hat{\xi}_1} \hat{\lambda}_2 - i_{\hat{\xi}_2} \hat{\lambda}_1) \\ \notag
&& +\kappa \Big({\cal L}_{\hat{\xi}_1}(i_{\hat{\chi}_2 \theta_0}h)-{\cal L}_{\hat{\xi}_2}(i_{\hat{\chi}_1 \theta_0}h) - \frac{1}{2}d(i_{\hat{\xi}_1} i_{\hat{\chi}_2 \theta_0} h -i_{\hat{\xi}_2} i_{\hat{\chi}_1 \theta_0} h) - i_{\hat{\chi}^c \theta_0} h \Big) + dB(\hat{\xi}_1, \hat{\xi}_2, .) \, .
\end{eqnarray}

\subsection{Courant algebroid interpretation of the $B$-frame}

We firstly note that the generalized currents $J_{\hat{B}}^M = (l^\mu , j_\mu)^T$ can be expressed by
\begin{equation} \label{eq:tildeX-def}
J_{\hat{B}}^M = (e^{\hat{B}})^M_{\ N} \hat{X}^N \, 
\end{equation}
where $\hat{X}^M$ is defined in (\ref{eq:hatX-def}). If we establish following relations between parameters
\begin{equation}
\hat{\Lambda}^M = (e^{\hat{B}})^M_{\ N} \Lambda^N \, , 
\end{equation}
we can recognize that $e^{\hat{B}}$ defines an isomorphism between $B$-twisted Courant bracket in T-dual picture and its non-twisted version
\begin{equation} \label{eq:hatCB-hatC}
[\hat{\Lambda}_1, \hat{\Lambda}_2]_{\hat{C}_{B}}= e^{\hat{B}}[e^{-\hat{B}} \hat{\Lambda}_1, e^{-\hat{B}}\hat{\Lambda}_2]_{\hat{C}} \, .
\end{equation}

The definition of the T-dual \(B\)-frame charge \({}^\star{\cal Q}^{B}_{\hat\Lambda}\) (\ref{eq:GB-star-dual}) is obtained from that of the initial \(\theta\)-frame charge \({\cal Q}^{\hat\theta}_{\Lambda}\) (\ref{eq:Gthdef}) by formally exchanging the initial coordinates and momenta with their T-dual counterparts and replacing the twisting bivector \(\kappa\theta^{\mu\nu}\) with the two-form \(2B_{\mu\nu}\). These replacements are accompanied by the exchange of vector and one-form parameters and the replacement of their argument \(x\) by \(V\). The resulting dual charge algebra closes on the weakly curved generalization of the \(B\)-twisted Courant bracket. Using (\ref{eq:Gth-Gth}), (\ref{eq:hatC-C}) and (\ref{eq:hatCB-hatC}), we obtain the isomorphism between the weakly curved generalization of the $B$-twisted Courant bracket and $\theta$-twisted Courant bracket 
\begin{equation}
[\hat{\Lambda}_1, \hat{\Lambda}_2]_{\hat{C}_{B}}= F_{\hat{B}}^{-1}[ F_{\hat{B}}\hat{\Lambda}_1, F_{\hat{B}}\hat{\Lambda}_2]_{{\cal C}_{\theta}} \, , \quad F_{\hat{B}} = e^{\hat{\theta}}F e^{-\hat{B}} \, ,
\end{equation}
where $F$ is isomorphism obtained in (\ref{eq:F-def}).
The appropriate symmetric pairing is as easily obtained
\begin{equation}
\eta_{F_{\hat{B}}} =  F_{\hat{B}}^T \eta  F_{\hat{B}} = (e^{-\hat{B}})^T \eta_F e^{-\hat{B}}\, , \quad (\eta_{F_{\hat{B}}})_{MN} = \begin{pmatrix}
2\kappa (b\theta_0 h - h \theta_0 b)_{\mu \nu} & \delta_\mu^\nu + \kappa h_{\mu \rho} \theta_0^{\rho \nu} \\ 
\delta^\mu_\nu + \kappa\theta_0^{\mu \rho}h_{\rho \nu} & 0 
\end{pmatrix} \, .
\end{equation}
and anchor is $\rho_B = \rho_{\theta} \circ F_{\hat{B}} = \rho_0$.

\section{The $\theta$-frame of the T-dual theory}
\cleq 

We will define the noncanonical currents in $\theta$-frame by 
\begin{equation}
{}^\star J_{\hat{\theta}}^M = ( e^{\hat{\theta}})_{\ N}^{M}\ {}^\star  X^N = \begin{pmatrix}
{}^\star \pi^\mu + \kappa^2 \theta^{\mu \nu}\big[V[y]\big] y_\nu^{\prime} \\
\kappa y_{\mu}^{\prime} \, ,
\end{pmatrix} \equiv \begin{pmatrix}
{}^\star k^\mu \\
\kappa y_{\mu}^{\prime}
\end{pmatrix} \, .
\end{equation}
We are interested in obtaining the fluxes that appear in the algebra of generalized currents ${}^\star J_{\hat{\theta}}^M$. Let us begin by defining ${}^\star J_{\hat{\theta}}^M \cong J_{\hat{\theta}}^M \, , {}^\star k^\mu  \cong k^\mu \, ,  \ \kappa y^{\prime}_{\mu} \cong i_\mu $, where
\begin{eqnarray} \label{eq:k-def}
k^\mu = \kappa x^{\prime \mu} + \kappa \theta^{\mu \nu} (x) \pi_\nu  \, ,
\end{eqnarray}

and $i_\mu$ is defined in (\ref{eq:aux-cur0}).

For future convenience, let us also express the canonical variables in terms of the currents\begin{subequations}
\begin{eqnarray} \label{eq:xpi-jk}
\kappa x^{\prime \mu} &=& (\delta^\mu_\nu - \kappa(\theta_0 h)^\mu_{\ \nu})\ k^\nu -\kappa (\delta^\mu_\rho - \kappa (\theta_0 h)^\mu_{\ \rho})\ \theta^{\rho \nu} i_\nu \, , \\ \label{eq:xpi-jk2}
\pi_\mu &=& (\delta^\nu_\mu - \kappa (h \theta_0)^{\ \nu}_{\mu})\ i_\nu +  h_{\mu \nu} k^\nu \, ,
\end{eqnarray}
\end{subequations}
where we disregarded terms in quadratic order of $h$. Now we obtain the algebra
\begin{eqnarray} \label{eq:kstar-kstar}
\{ k^\mu  (\sigma), k^\nu (\bar{\sigma}) \} &=& -\kappa Q_\rho^{\ \mu \nu} k^\rho \delta(\sigma-\bar{\sigma}) - \kappa^2 R^{\mu \nu \rho} \pi_\rho \delta(\sigma-\bar{\sigma}) \\ \notag
&=& -\kappa Q_\rho^{\ \mu \nu} k^\rho \delta(\sigma-\bar{\sigma})- \kappa^2 R^{\mu \nu \rho} i_\rho \delta(\sigma-\bar{\sigma}) \, ,
\end{eqnarray}
where we defined the fluxes by
\begin{eqnarray} \label{eq:QR-def}
Q_\rho^{\ \mu \nu} = \partial_\rho \theta^{\mu \nu} \, , \quad R^{\mu \nu \rho} = \theta^{\mu \sigma} \partial_\sigma \theta^{\nu \rho}+\theta^{\nu \sigma} \partial_\sigma  \theta^{\rho \mu}+\theta^{\rho \sigma} \partial_\sigma  \theta^{\mu \nu} \, .
\end{eqnarray}
In the last row of (\ref{eq:kstar-kstar}), we used that in the weakly curved approximation $R^{\mu \nu \rho} = {\cal O}(h)$ and $\pi_\mu = i_\mu + {\cal O}(h)$.

On the other hand, we have
\begin{eqnarray} \label{eq:kstar-jstar}
\{ i_\mu (\sigma), k^{\nu} (\bar{\sigma}) \} &=& \kappa (\delta_\mu^\nu + \kappa (\theta_0 h)^\nu_{\ \mu}) \delta^{\prime}(\sigma-\bar{\sigma}) \\ \notag
&&- \Big(\kappa Q_\mu^{\ \nu \rho}+ \frac{\kappa^2}{3} \theta_0^{\nu \sigma} \theta_0^{\rho \tau} B_{\sigma \mu \tau} \Big) i_\rho \delta(\sigma-\bar{\sigma})  -\frac{\kappa}{3} \theta_0^{\nu \sigma} B_{\sigma \mu \rho} k^\rho \delta(\sigma-\bar{\sigma}) \, .
\end{eqnarray}
Lastly, we can just rewrite (\ref{eq:NC1}) with the help of (\ref{eq:xpi-jk}) in order to obtain
\begin{eqnarray}
\{ i_\mu (\sigma), i_\nu (\bar{\sigma}) \} =  B_{\mu \nu \rho} k^\rho \delta(\sigma-\bar{\sigma}) -\kappa B_{\mu \nu \sigma} \theta_0^{\sigma \rho } i_\rho \delta(\sigma-\bar{\sigma}) \, .
\end{eqnarray}

\subsection{Generalized charge algebra and the $\theta$-twisted bracket}

In the same way as in previous chapter, we define
\begin{eqnarray} \label{eq:Gth-star-dual}
&& {}^{\star} {\cal Q}^{\theta}_{\hat{\Lambda}} = \int_0^{2\pi} d\sigma\  \hat{\Lambda}^M \eta_{MN}\ {}^\star J_{\hat{\theta}}^N \, \\ \notag
&& {}^{\star} {\cal Q}^{\theta}_{\hat{\Lambda}} \cong \int_0^{2\pi} d\sigma \Big( \hat{\xi}^\mu (x) i_{\mu} + \hat{\lambda}_\mu (x) k^\mu \Big) \, ,
\end{eqnarray}
where $\hat{\Lambda}^M$ has the same form as in (\ref{eq:Lstar-def}). In addition to the relations (\ref{eq:auxPB1}) - (\ref{eq:auxPB3}), we will require
\begin{equation}\label{eq:auxPB5}
\begin{aligned}
\left\{
V^\mu(\sigma),
{}^\star k^\nu(\bar{\sigma})
\right\}
&\cong
\left\{
x^\mu(\sigma) ,
\kappa x^{\prime\nu}(\bar{\sigma})
+
\kappa \theta^{\nu\rho} (\bar{\sigma})
\pi_\rho (\bar{\sigma})
\right\}=-\kappa \theta^{\mu \nu} \delta(\sigma-\bar{\sigma})  \, .
\end{aligned}
\end{equation}
First, we have
\begin{eqnarray} \label{eq:aux31}
\{ \hat{\lambda}_{1\mu} (\sigma) k^\mu (\sigma), \hat{\lambda}_{2\nu} (\bar{\sigma}) k^\nu (\bar{\sigma}) \} &=& -\kappa \Big[(\hat{\lambda}_1 \theta)^\nu \partial_\nu \hat{\lambda}_{2\mu} - (\hat{\lambda}_2 \theta)^\nu \partial_\nu \hat{\lambda}_{1\mu} \\ \notag
&& \qquad +  Q_\mu^{\ \nu \rho} \hat{\lambda}_{1\nu} \hat{\lambda}_{2\rho} \Big] k^\mu \delta(\sigma-\bar{\sigma}) \\ \notag
&&-\kappa^2 R^{\mu \nu \rho} \hat{\lambda}_{1\nu} \hat{\lambda}_{2\rho} i_\mu \delta(\sigma-\bar{\sigma}) \, .
\end{eqnarray}
Second,we obtain
\begin{eqnarray} \label{eq:aux33}
\{ \hat{\xi}_1^\mu (\sigma) i_\mu (\sigma), \hat{\xi}_2^\nu (\bar{\sigma}) i_\nu (\bar{\sigma}) \} &=& -(\hat{\xi}_1^\nu \partial_\nu \hat{\xi}_2^\mu - \hat{\xi}_2^\nu \partial_\nu \hat{\xi}_1^\mu - \kappa \theta^{\mu \nu} B_{\nu \rho \sigma} \hat{\xi}_1^\rho \hat{\xi}_2^\sigma) i_\mu \delta(\sigma-\bar{\sigma}) \\ \notag
&& + B_{\mu \nu \rho} \hat{\xi}_1^\nu \hat{\xi}_2^\rho k^\mu \delta(\sigma-\bar{\sigma})\, . 
\end{eqnarray}
The mixed term produces
\begin{eqnarray} \label{eq:aux321}
\{ \hat{\xi}_1^{\mu} (\sigma) i_\mu  (\sigma), \hat{\lambda}_{2\nu} (\bar{\sigma}) k^\nu  (\bar{\sigma}) \} &=& {\cal A}(\sigma, \bar{\sigma}) \\ \notag
&&- \Big[\hat{\xi}_1^\nu \partial_\nu \hat{\lambda}_{2\mu} + \frac{\kappa}{3} (\hat{\lambda}_2 \theta_0)^\rho B_{\rho \nu \mu} \hat{\xi}_1^{\nu} \Big] k^\mu \delta(\sigma-\bar{\sigma}) \\ \notag
&& -\kappa \Big[- (\hat{\lambda}_2 \theta)^{\nu}\partial_\nu \hat{\xi}_1^\mu + Q_\nu^{\rho \mu } \hat{\lambda}_{2\rho} \hat{\xi}_1^\nu \\ \notag
&&\qquad +\frac{\kappa}{3}\hat{\lambda}_{2\rho} \theta_0^{\rho \sigma} \theta_0^{\mu \tau} B_{\sigma \nu \tau} \hat{\xi}_1^\nu \Big] i_\mu \delta(\sigma-\bar{\sigma}).
\end{eqnarray}
where ${\cal A}(\sigma, \bar{\sigma})$ is defined in (\ref{eq:aux1an}). Substituting (\ref{eq:xpi-jk}) into (\ref{eq:auxAN}), we obtain the anomalous part in terms of $i_\mu$ and $k^\mu$ currents
\begin{eqnarray} \label{eq:aux322}
{\cal A}(\sigma, \bar{\sigma}) &=& {\cal A}_0(\sigma, \bar{\sigma}) \\ \notag
&& + \Big[\hat{\xi}_1^\nu \partial_\mu\hat{\lambda}_{2\nu} - \frac{1}{2} \partial_\mu (\hat{\xi}_1 \hat{\lambda}_2) + \kappa \hat{\xi}_1^\rho (h\theta_0)_\rho^{\ \nu} \partial_\mu \hat{\lambda}_{2\nu} - \frac{\kappa}{2}\partial_\mu (\hat{\xi}_1 h \theta_0 \hat{\lambda}_2) \\ \notag
&& \quad -\kappa (\theta_0h)^\rho_{\ \mu} \hat{\xi}_1^\nu \partial_\rho \hat{\lambda}_{2\nu} + \frac{\kappa}{2} (\theta_0 h)^\rho_{\ \mu} \partial_\rho (\hat{\xi}_1 \hat{\lambda}_2) \Big] k^\mu \delta(\sigma-\bar{\sigma}) \\ \notag
&&- \kappa \Big[\hat{\xi}_1^\nu \partial_\rho \hat{\lambda}_{2\nu} - \frac{1}{2} \partial_\rho (\hat{\xi}_1 \hat{\lambda}_2) + \kappa \hat{\xi}_1^\sigma (h\theta_0)_\sigma^{\ \nu} \partial_\rho \hat{\lambda}_{2\nu} - \frac{\kappa}{2}\partial_\rho (\hat{\xi}_1 h \theta_0 \hat{\lambda}_2) \\ \notag
&& \quad -\kappa (\theta_0h)^\sigma_{\ \rho} \hat{\xi}_1^\nu \partial_\sigma \hat{\lambda}_{2\nu} + \frac{\kappa}{2} (\theta_0 h)^\sigma_{\ \rho} \partial_\sigma (\hat{\xi}_1 \hat{\lambda}_2) \Big] \theta^{\rho \mu} i_\mu \delta(\sigma-\bar{\sigma}) \, .
\end{eqnarray}
where ${\cal A}_0(\sigma, \bar{\sigma})$ is pure anomaly (\ref{eq:auxA0}) which disappears after integration over both $\sigma$ and $\bar{\sigma}$. Substituting (\ref{eq:aux322}) into (\ref{eq:aux321}) and using the explicit expression for $Q$-flux (\ref{eq:QR-def}), we obtain 
\begin{eqnarray} \label{eq:aux323}
\{ \hat{\xi}_1^{\mu} (\sigma) i_\mu  (\sigma), \hat{\lambda}_{2\nu} (\bar{\sigma}) k^\nu  (\bar{\sigma})\} &=& {\cal A}_0(\sigma, \bar{\sigma}) \\ \notag
&&- \Big[\hat{\xi}_1^\nu (\partial_\nu \hat{\lambda}_{2\mu}-\partial_\mu \hat{\lambda}_{2\nu}) + \frac{1}{2}\partial_\mu (\hat{\xi}_1 \hat{\lambda}_2) + \frac{\kappa}{3} (\hat{\lambda}_2 \theta_0)^\rho B_{\rho \nu \mu} \hat{\xi}_1^{\nu}  \\ \notag
&&\quad- \kappa \hat{\xi}_1^\rho (h\theta_0)_\rho^{\ \nu} \partial_\mu \hat{\lambda}_{2\nu}+ \frac{\kappa}{2}\partial_\mu (\hat{\xi}_1 h \theta_0 \hat{\lambda}_2)  \\ \notag
&&\quad+\kappa (\theta_0h)^\rho_{\ \mu} \hat{\xi}_1^\nu \partial_\rho \hat{\lambda}_{2\nu} - \frac{\kappa}{2} (\theta_0 h)^\rho_{\ \mu} \partial_\rho (\hat{\xi}_1 \hat{\lambda}_2) \Big] k^\mu \delta(\sigma-\bar{\sigma}) \\ \notag
&&-\kappa \Big[\hat{\xi}_1^\nu \partial_\nu (\hat{\lambda}_2\theta)^\mu - (\hat{\lambda}_2\theta)^\nu \partial_\nu \hat{\xi}_1^\mu  +\frac{\kappa}{3}\hat{\lambda}_{2\rho} \theta_0^{\rho \sigma} \theta_0^{\mu \tau} B_{\sigma \nu \tau} \hat{\xi}_1^\nu \\ \notag
&&\quad+ \theta^{ \mu \rho}\Big( \hat{\xi}_1^\nu (\partial_\nu \hat{\lambda}_{2\rho}-\partial_\rho \hat{\lambda}_{2\nu}) + \frac{1}{2}\partial_\rho (\hat{\xi}_1 \hat{\lambda}_2)  \\ \notag
&&\qquad  -\kappa \hat{\xi}_1^\sigma (h\theta_0)_\sigma^{\ \nu} \partial_\rho \hat{\lambda}_{2\nu} + \frac{\kappa}{2}\partial_\rho (\hat{\xi}_1 h \theta_0 \hat{\lambda}_2) \\ \notag
&& \qquad+\kappa (\theta_0h)^\sigma_{\ \rho} \hat{\xi}_1^\nu \partial_\sigma \hat{\lambda}_{2\nu} - \frac{\kappa}{2} (\theta_0 h)^\sigma_{\ \rho} \partial_\sigma (\hat{\xi}_1 \hat{\lambda}_2) 
\Big) \Big] i_\mu \delta(\sigma-\bar{\sigma}) \, .
\end{eqnarray}
In order to write results in consistent way with previous chapters, we will apply the chain rule on the expressions of the following type 
\begin{eqnarray} \label{eq:aux324}
\partial_\rho (\hat{\xi}_1 h \theta_0 \hat{\lambda}_2) &=&  -\frac{1}{3}B_{\rho \sigma \tau} \hat{\xi}_1^\sigma (\hat{\lambda}_{2} \theta_0)^{\tau}+(\theta_0 h)_{\ \sigma}^{\nu} \partial_\rho (\hat{\xi}_1^\sigma \hat{\lambda}_{2\nu}) \, ,
\end{eqnarray}
which substituting to (\ref{eq:aux323}) results in
\begin{eqnarray} \label{eq:aux32}
\{ \hat{\xi}_1^{\mu} (\sigma) i_\mu  (\sigma), \hat{\lambda}_{2\nu} (\bar{\sigma}) k^\nu  (\bar{\sigma})\} &=& {\cal A}_0(\sigma, \bar{\sigma}) \\ \notag
&&- \Big[\hat{\xi}_1^\nu (\partial_\nu \hat{\lambda}_{2\mu}-\partial_\mu \hat{\lambda}_{2\nu}) + \frac{1}{2}\partial_\mu (\hat{\xi}_1 \hat{\lambda}_2) + \frac{\kappa}{2} (\hat{\lambda}_2 \theta_0)^\rho B_{\rho \nu \mu} \hat{\xi}_1^{\nu}  \\ \notag
&&\quad- \kappa (\theta_0 h)_{\ \rho}^{\nu} ( \hat{\xi}_1^\rho \partial_\mu \hat{\lambda}_{2\nu}- \frac{1}{2}\partial_\mu (\hat{\xi}_1^\rho \hat{\lambda}_{2\nu}))  \\ \notag
&&\quad+\kappa (\theta_0h)^\rho_{\ \mu} (\hat{\xi}_1^\nu \partial_\rho \hat{\lambda}_{2\nu} - \frac{1}{2}  \partial_\rho (\hat{\xi}_1 \hat{\lambda}_2)) \Big] k^\mu \delta(\sigma-\bar{\sigma}) \\ \notag
&&-\kappa \Big[\hat{\xi}_1^\nu \partial_\nu (\hat{\lambda}_2\theta)^\mu - (\hat{\lambda}_2\theta)^\nu \partial_\nu \hat{\xi}_1^\mu  +\frac{\kappa}{2}\hat{\lambda}_{2\rho} \theta_0^{\rho \sigma} \theta_0^{\mu \tau} B_{\sigma \nu \tau} \hat{\xi}_1^\nu \\ \notag
&&\quad+ \theta^{ \mu \rho}\Big( \hat{\xi}_1^\nu (\partial_\nu \hat{\lambda}_{2\rho}-\partial_\rho \hat{\lambda}_{2\nu}) + \frac{1}{2}\partial_\rho (\hat{\xi}_1 \hat{\lambda}_2)  \\ \notag
&&\qquad  -\kappa (\theta_0 h)_{\ \sigma}^{\nu} (\hat{\xi}_1^\sigma \partial_\rho \hat{\lambda}_{2\nu} - \frac{1}{2}\partial_\rho (\hat{\xi}_1^\sigma \hat{\lambda}_{2\nu})) \\ \notag
&& \qquad+\kappa (\theta_0h)^\sigma_{\ \rho} (\hat{\xi}_1^\nu \partial_\sigma \hat{\lambda}_{2\nu} - \frac{1}{2} \partial_\sigma (\hat{\xi}_1 \hat{\lambda}_2)) 
\Big) \Big] i_\mu \delta(\sigma-\bar{\sigma}) \, .
\end{eqnarray}

Using (\ref{eq:aux31}), (\ref{eq:aux33}) and (\ref{eq:aux32}), we obtain
\begin{equation} \label{eq:Gstar-aux2}
\{ {}^{\star} {\cal Q}^{\theta}_{\hat{\Lambda}_1} , {}^{\star} {\cal Q}^{\theta}_{\hat{\Lambda}_2}  \} =- {}^{\star} {\cal Q}^{\theta}_{[\hat{\Lambda}_1, \hat{\Lambda}_2]_{\hat{C}_{\theta}}} \, , 
\end{equation}
where the algebra closes on weakly curved $\theta$-twisted generalisation of the Courant bracket. If we assume notation $ {}^{\star}{\cal Q}^{\theta}_{[\hat{\Lambda}_1, \hat{\Lambda}_2]_{\hat{C}_{\theta}}} \cong \int d\sigma \Big( \hat{\xi}^\mu (x) i_\mu + \hat{\lambda}_\mu(x) k^\mu \Big) $, the resulting parameters are given by 
\begin{eqnarray} \label{eq:Courant-dual30}
\hat{\xi}^\mu &=& \hat{\xi}_1^\nu \partial_\nu \hat{\xi}_2^\mu - \hat{\xi}_2^\nu \partial_\nu \hat{\xi}_1^\mu\\ \notag
&&+\kappa \hat{\xi}_1^\nu \partial_\nu (\hat{\lambda}_2\theta)^\mu -\kappa  (\hat{\lambda}_2\theta)^\nu \partial_\nu \hat{\xi}_1^\mu-\kappa \hat{\xi}_2^\nu \partial_\nu (\hat{\lambda}_1\theta)^\mu + \kappa (\hat{\lambda}_1\theta)^\nu \partial_\nu \hat{\xi}_2^\mu \\ \notag
&& +\kappa \theta^{ \mu \rho}\Big( \hat{\xi}_1^\nu (\partial_\nu \hat{\lambda}_{2\rho}-\partial_\rho \hat{\lambda}_{2\nu})-\hat{\xi}_2^\nu (\partial_\nu \hat{\lambda}_{1\rho}-\partial_\rho \hat{\lambda}_{1\nu}) + \frac{1}{2}\partial_\rho (\hat{\xi}_1 \hat{\lambda}_2 - \hat{\xi}_2 \hat{\lambda}_1) \Big) \\ \notag
&&  +\kappa \theta^{\mu \rho} \Big(-\kappa(\theta_0 h)_{\ \sigma}^{\nu} (\hat{\xi}_1^\sigma \partial_\rho \hat{\lambda}_{2\nu} -\hat{\xi}_2^\sigma  \partial_\rho \hat{\lambda}_{1\nu}-\frac{1}{2} \partial_\rho (\hat{\xi}_1^\sigma \hat{\lambda}_{2\nu}-\hat{\xi}_2^\sigma \hat{\lambda}_{1\nu})) \\ \notag
&&\qquad +\kappa (\theta_0h)^\sigma_{\ \rho} (\hat{\xi}_1^\nu \partial_\sigma \hat{\lambda}_{2\nu}- \hat{\xi}_2^\nu \partial_\sigma \hat{\lambda}_{1\nu}- \frac{1}{2} \partial_\sigma (\hat{\xi}_1 \hat{\lambda}_2-\hat{\xi}_2 \hat{\lambda}_1)) \Big) \\ \notag
&& - \kappa \theta^{\mu \nu} B_{\nu \rho \sigma} \hat{\xi}_1^\rho \hat{\xi}_2^\sigma +\frac{\kappa^2}{2}\theta_0^{\mu \rho} (\hat{\lambda}_2 \theta_0)^\sigma B_{\sigma \nu \rho} \hat{\xi}_1^{\nu} -\frac{\kappa^2}{2}\theta_0^{\mu \rho} (\hat{\lambda}_1 \theta_0)^\sigma B_{\sigma \nu \rho}  \hat{\xi}_2^{\nu} + \kappa^2 R^{\mu \nu \rho} \hat{\lambda}_{1\nu} \hat{\lambda}_{2\rho} \, ,
\end{eqnarray}
and
\begin{eqnarray} \label{eq:Courant-dual31}
\hat{\lambda}_{\mu}
&=&  \hat{\xi}_1^\nu (\partial_\nu \hat{\lambda}_{2\mu}-\partial_\mu \hat{\lambda}_{2\nu})-\hat{\xi}_2^\nu (\partial_\nu \hat{\lambda}_{1\mu}-\partial_\mu \hat{\lambda}_{1\nu}) + \frac{1}{2}\partial_\mu (\hat{\xi}_1 \hat{\lambda}_2- \hat{\xi}_2 \hat{\lambda}_1)  \\ \notag
&&+\kappa ((\hat{\lambda}_1 \theta)^\nu \partial_\nu \hat{\lambda}_{2\mu} - (\hat{\lambda}_2 \theta)^\nu \partial_\nu \hat{\lambda}_{1\mu} +  Q_\mu^{\ \nu \rho} \hat{\lambda}_{1\nu} \hat{\lambda}_{2\rho})\\ \notag
&& +\kappa (\theta_0h)^\rho_{\ \mu} (\hat{\xi}_1^\nu \partial_\rho\hat{\lambda}_{2\nu}- \hat{\xi}_2^\nu \partial_\rho \hat{\lambda}_{1\nu}- \frac{1}{2}  \partial_\rho (\hat{\xi}_1 \hat{\lambda}_2-\hat{\xi}_2 \hat{\lambda}_1))  \\ \notag
&&  -\kappa(\theta_0 h)_{\ \rho}^{\nu} (\hat{\xi}_1^\rho \partial_\mu \hat{\lambda}_{2\nu} -\hat{\xi}_2^\rho \partial_\mu \hat{\lambda}_{1\nu}-\frac{1}{2}\partial_\mu (\hat{\xi}_1^\rho \hat{\lambda}_{2\nu} - \hat{\xi}_2^\rho \hat{\lambda}_{1\nu})) \\ \notag
&&-B_{\mu \nu \rho} \hat{\xi}_1^\nu \hat{\xi}_2^\rho+\frac{\kappa}{2} (\hat{\lambda}_2 \theta_0)^\rho B_{\rho \nu \mu} \hat{\xi}_1^{\nu}-\frac{\kappa}{2} (\hat{\lambda}_1 \theta_0)^\rho B_{\rho \nu \mu} \hat{\xi}_2^{\nu} \, .
\end{eqnarray}

The non-coordinate expressions can easily be deduced from (\ref{eq:thcourant-nc}) and (\ref{eq:gen-courant-nc})
\begin{eqnarray} \label{eq:genthe-courant-nc}
\hat{\xi} &=& [\hat{\xi}_1, \hat{\xi}_2]_L - \kappa [\hat{\xi}_2, \hat{\lambda}_1 \theta]_L +  \kappa [\hat{\xi}_1, \hat{\lambda}_2 \theta]_L + \frac{\kappa^2}{2}[\theta, \theta]_S (\hat{\lambda}_1, \hat{\lambda}_2, .)\\ \notag
&& -\kappa \theta \Big( {\cal{L}}_{\hat{\xi}_1} \hat{\lambda}_2 -{\cal{L}}_{\hat{\xi}_2} \hat{\lambda}_1 - \frac{1}{2} d (i_{\hat{\xi}_1} \hat{\lambda}_2 - i_{\hat{\xi}_2} \hat{\lambda}_1 ) \Big) \\ \notag
&&- \kappa^2 \theta\Big({\cal L}_{\hat{\xi}_1}(i_{\hat{\lambda}_2 \theta_0}h)-{\cal L}_{\hat{\xi}_2}(i_{\hat{\lambda}_1 \theta_0}h) - \frac{1}{2}d(i_{\hat{\xi}_1} i_{\hat{\lambda}_2 \theta_0} h -i_{\hat{\xi}_2} i_{\hat{\lambda}_1 \theta_0} h) - i_{\hat{\lambda}^c \theta_0} h \Big)\\ \notag
&&+\kappa \theta(d B(\hat{\xi}_1, \hat{\xi}_2, .))+\kappa^2 \theta(dB(\hat{\xi}_1, \hat{\lambda}_2 \theta_0, .))-\kappa^2 \theta(dB(\hat{\xi}_2, \hat{\lambda}_1 \theta_0, .)) \\ \notag
\hat{\lambda} &=& {\cal{L}}_{\hat{\xi}_1} \hat{\lambda}_2 - {\cal{L}}_{\hat{\xi}_2} \hat{\lambda}_1 - \frac{1}{2}d (i_{\hat{\xi}_1} \hat{\lambda}_2 - i_{\hat{\xi}_2} \hat{\lambda}_1) + \kappa [\hat{\lambda}_1 ,\hat{\lambda}_2]_{ \theta}\\ \notag
&& +\kappa \Big({\cal L}_{\hat{\xi}_1}(i_{\hat{\lambda}_2 \theta_0}h)-{\cal L}_{\hat{\xi}_2}(i_{\hat{\lambda}_1 \theta_0}h) - \frac{1}{2}d(i_{\hat{\xi}_1} i_{\hat{\lambda}_2 \theta_0} h -i_{\hat{\xi}_2} i_{\hat{\lambda}_1 \theta_0} h) - i_{\hat{\lambda}^c \theta_0} h \Big)\\ \notag
&& - dB(\hat{\xi}_1, \hat{\xi}_2, .) - \kappa dB(\hat{\xi}_1, \hat{\lambda}_2 \theta_0, .)+ \kappa dB(\hat{\xi}_2, \hat{\lambda}_1 \theta_0, .) \, .
\end{eqnarray}

\subsection{Courant algebroid interpretation of the $\theta$-frame}

For completeness, let us also demonstrate the various isomorphisms for the weakly curved generalization of the $\theta$-twisted Courant bracket.
We firstly note that the generalized currents $J_{\hat{\theta}}^M = (k^\mu , i_\mu)^T$ can be expressed by
\begin{equation} \label{eq:J-th-def}
J_{\hat{\theta}}^M = (e^{\hat{\theta}})^M_{\ N} \hat{X}^N \, 
\end{equation}
where $\hat{X}^M$ is defined in (\ref{eq:hatX-def}). If we establish following relations between parameters
\begin{equation}
\hat{\Lambda}^M = (e^{\hat{\theta}})^M_{\ N} \Lambda^N \, , 
\end{equation}
we can recognize that $e^{\hat{\theta}}$ defines an isomorphism between $\theta$-twisted Courant bracket in T-dual picture and its non-twisted version
\begin{equation} \label{eq:hatCth-hatC}
[\hat{\Lambda}_1, \hat{\Lambda}_2]_{\hat{C}_{\theta}}= e^{\hat{\theta}}[e^{-\hat{\theta}} \hat{\Lambda}_1, e^{-\hat{\theta}}\hat{\Lambda}_2]_{\hat{C}} \, .
\end{equation}
Similarly, the definition of the T-dual \(\theta\)-frame charge \({}^\star{\cal Q}^{\theta}_{\hat\Lambda}\) (\ref{eq:Gth-star-dual}) is obtained from that of the initial \(B\)-frame charge \({\cal Q}^{\hat B}_{\Lambda}\) (\ref{eq:GBdef}) by formally exchanging the initial coordinates and momenta with their T-dual counterparts and replacing the Kalb-Ramond field \(2B_{\mu\nu}\) with its T-dual non-commutativity parameter \(\kappa\theta^{\mu\nu}\). The vector and one-form parameters are exchanged accordingly, and their argument \(x\) is replaced by \(V\). The resulting dual charge algebra closes on the weakly curved generalization of the \(\theta\)-twisted Courant bracket. The isomorphism between these brackets is obtained from (\ref{eq:GBGB-GB}), (\ref{eq:hatC-C}) and (\ref{eq:hatCth-hatC}), which gives relation
\begin{equation}
[\hat{\Lambda}_1, \hat{\Lambda}_2]_{\hat{C}_{\theta}} = F_{\hat{\theta}}^{-1} [F_{\hat{\theta}} \hat{\Lambda}_1, F_{\hat{\theta}}\hat{\Lambda}_2]_{{\cal C}_{B}} \, , \quad F_{\hat{\theta}} = e^{\hat{B}} F e^{-\hat{\theta}} \, ,
\end{equation}
where $F$ is defined in (\ref{eq:F-def}). The pairing is given by
\begin{equation}
\eta_{F_{\hat{\theta}}} =  F_{\hat{\theta}}^T \eta  F_{\hat{\theta}} = (e^{-\hat{\theta}})^T \eta_F e^{-\hat{\theta}} = \eta_F \, ,
\end{equation}
which is the same pairing as in (\ref{eq:eta-F}), while the anchor is $\rho_{\hat{\theta}} =  \rho_0 \circ F_{\hat{\theta}} $.

\section{Conclusions}
\cleq

In this paper, we studied a closed bosonic string moving in a weakly curved background, characterized by a constant metric and a Kalb-Ramond field with linear coordinate dependence. All calculations were performed to first order in the constant Kalb-Ramond field strength. In the T-dual description, both background fields depend on the line integral $V(y_\mu,\tilde y_\mu)$, making the theory nonlocal in the T-dual coordinates. The T-duality relations contain field strength dependent corrections to the usual exchange of momenta and coordinate $\sigma$-derivatives. The Poisson brackets inherited from the initial theory are noncanonical in the chosen T-dual variables, and the dual coordinates are noncommutative. We investigated generalized currents and their charges in this T-dual description, identified the Courant algebroid structures governing their algebras and established their relations to the well-known Courant algebroids obtained in the initial theory.

Firstly, we considered generalized currents expressed in terms of the T-dual phase space variables, with parameters depending on the same line integral as the T-dual background fields. Using T-duality induced Poisson brackets between variables in the T-dual phase space, we obtained the full Poisson bracket algebra of generalized charges of these currents. The charge algebra yields a weakly curved generalization of the Courant bracket, which we obtained in both coordinate and coordinate-free form. The additional terms originate from the field strength corrections to the T-duality relations. We established an isomorphism with the standard Courant algebroid by identifying the corresponding transformation of the bracket, anchor, and pairing. The transporting map is generally not \(O(D,D)\) with respect to the standard pairing. The Courant algebroid isomorphism therefore requires transporting the pairing together with the bracket and anchor.

When T-duality is a simple canonical transformation, Courant bracket is invariant under exchange of phase space variables with their T-dual analogs. In case of weakly curved background this is no longer the case and the Courant bracket becomes its weakly curved generalization. 

Secondly, we considered the $B$-frame, obtained by applying the $B$-shift to the generalized currents of the preceding construction. We obtained a $B$-twisted weakly curved generalization of the Courant bracket, as well as its corresponding algebroid. Under the formal replacement of the initial variables by their T-dual counterparts, accompanied by the exchanges of twisting fields and parameters described above, the initial $\theta$-frame generalized charge maps to this T-dual $B$-frame generalized charge. We established an explicit isomorphism between the corresponding Courant algebroid structures.

Finally, we considered the $\theta$-frame. Its charge algebra yields a $\theta$-twisted weakly curved generalization of the Courant bracket. The analogous formal replacements map the initial $B$-frame currents to these T-dual $\theta$-frame currents. We likewise established an explicit isomorphism between the corresponding Courant algebroids.

These results show that the algebra of generalized charges in this nonlocal T-dual theory admits a Courant algebroid description. Although the charges depend nonlocally on the dual fields, their algebra is related to the familiar Courant and twisted Courant structures by explicit isomorphisms. The additional perturbative contributions to the brackets, together with the transformed anchors and pairings, describe how these structures are realized in the T-dual variables.

This provides a connection with the interpretation of T-duality through Courant algebroid isomorphisms developed by Cavalcanti and Gualtieri \cite{Tdualiso}. Their construction relates the algebroids of invariant sections associated with T-dual torus bundles carrying $H$-flux. Our results parallel this principle at the level of generalized charge algebras for a T-dual pair whose dual description is nonlocal. Thus, within the weakly curved approximation considered, the line integral dependence of the dual theory remains compatible with an isomorphism of Courant algebroid structures.

Further work could investigate how more general background dependence determines corrections to the generalized current algebras and their associated brackets, anchors, and pairings. A systematic understanding of these corrections could clarify the conditions under which a Courant algebroid description persists and whether this description can guide T-dualization beyond the weakly curved approximation.

\vspace{2 cm}
\textbf{Acknowledgements}
\vspace{.5 cm}

This work is supported in part by the Serbian Ministry of Science, and the Institute of Physics Belgrade.

In this work Large Language Models (ChatGPT 6.1 Sol, Claude 5.5 Sonnet) were used to verify calculations and to assist with language editing. All scientific results were independently verified by the author, who also takes full responsibility for their interpretation.

\appendix

\section{Line integral in the first order}\label{app:v}

 In the first order, the line integral is given by \cite{T-dual-2nd}
\begin{equation}
\begin{aligned}
V_1^\mu(y)
={}&
-\frac{\kappa}{2}
\left(
\Theta_{0+}^{\mu\nu}
+
\Theta_{0-}^{\mu\nu}
\right)y_\nu^{(1)}
-\frac{\kappa}{2}
\left(
\Theta_{0+}^{\mu\nu}
-
\Theta_{0-}^{\mu\nu}
\right)\widetilde{y}_\nu^{(1)}
\\[2mm]
&-\frac{\kappa^3}{2}
\left(
\Theta_{0+}^{\mu\nu}
+
\Theta_{0-}^{\mu\nu}
\right)
B_{\nu\rho\sigma}
\Bigl[
    \Theta_{0+}^{\sigma\tau}
    \Theta_{0+}^{\rho\varepsilon}
    M_{+\tau\varepsilon}(y)
\\
&\hspace{42mm}
    +
    \Theta_{0-}^{\sigma\tau}
    \Theta_{0-}^{\rho\varepsilon}
    M_{-\tau\varepsilon}(y)
    +
    \Theta_{0-}^{\sigma\tau}
    \Theta_{0+}^{\rho\varepsilon}
    \widetilde{
        \left(
        y_{-\tau}^{(0)}
        y_{+\varepsilon}^{(0)}
        \right)
    }
\Bigr]
\\[2mm]
&-\frac{\kappa^3}{2}
\left(
\Theta_{0+}^{\mu\nu}
-
\Theta_{0-}^{\mu\nu}
\right)
B_{\nu\rho\sigma}
\Bigl[
    \Theta_{0+}^{\sigma\tau}
    \Theta_{0+}^{\rho\varepsilon}
    M_{+\tau\varepsilon}(y)
\\
&\hspace{42mm}
    -
    \Theta_{0-}^{\sigma\tau}
    \Theta_{0-}^{\rho\varepsilon}
    M_{-\tau\varepsilon}(y)
    +
    \Theta_{0-}^{\sigma\tau}
    \Theta_{0+}^{\rho\varepsilon}
    y_{-\tau}^{(0)}
    y_{+\varepsilon}^{(0)}
\Bigr] \, ,
\end{aligned}
\label{eq:V1}
\end{equation}
where $\widetilde{(y_{-\tau}^{(0)}y_{+\varepsilon}^{(0)})}$ is defined as in (\ref{eq:tildey}), and
\begin{equation}
y_{\pm \mu}^{(0)} = \frac{1}{2}(y_\mu^{(0)} \pm \tilde{y}_\mu^{(0)})\, ,
\end{equation}
and
\begin{equation}
M_{\pm\mu\nu}(y)
\equiv
\frac{1}{2}
\int d\xi^\pm
\left(
y_{\pm\mu}^{(0)}\partial_\pm y_{\pm\nu}^{(0)}
-
y_{\pm\nu}^{(0)}\partial_\pm y_{\pm\mu}^{(0)}
\right) \, .
\end{equation}

\section{Courant algebroids and their isomorphisms}\label{app:courant}
\cleq

Let $E \to M$ be a vector bundle equipped with a nondegenerate symmetric bilinear form $\eta$, a skew-symmetric bracket $[\, ,\, ]$ on its smooth sections, and a smooth bundle map $\rho:E\to TM$, called the anchor. Define the differential operator ${\cal D}:C^\infty(M)\to\Gamma(E)$ by
\begin{equation} \label{eq:calDdef}
\eta({\cal D}f,\Lambda)
=
{\cal L}_{\rho(\Lambda)}f \, .
\end{equation}
The structure $\big(E,\eta,[\, ,\, ],\rho\big)$ is called a Courant algebroid if it satisfies the following compatibility conditions:
\begin{align}
\rho([\Lambda_1,\Lambda_2])
&=
[\rho(\Lambda_1),\rho(\Lambda_2)]_L \, ,
\label{eq:CAdef1}
\\
[\Lambda_1,f\Lambda_2]
&=
f[\Lambda_1,\Lambda_2]
+
\big({\cal L}_{\rho(\Lambda_1)}f\big)\Lambda_2
-
\frac12\eta(\Lambda_1,\Lambda_2){\cal D}f \, ,
\label{eq:CAdef2}
\\
{\cal L}_{\rho(\Lambda_1)}\eta(\Lambda_2,\Lambda_3)
&=
\eta\left(
[\Lambda_1,\Lambda_2]
+\frac12{\cal D}\eta(\Lambda_1,\Lambda_2),
\Lambda_3
\right)
\nonumber\\
&\quad+
\eta\left(
\Lambda_2,
[\Lambda_1,\Lambda_3]
+\frac12{\cal D}\eta(\Lambda_1,\Lambda_3)
\right) \, ,
\label{eq:CAdef3}
\\
\eta({\cal D}f,{\cal D}g)
&=0 \, ,
\label{eq:CAdef4}
\\
\operatorname{Jac}(\Lambda_1,\Lambda_2,\Lambda_3)
&=
{\cal D}\operatorname{Nij}(\Lambda_1,\Lambda_2,\Lambda_3) \, ,
\label{eq:CAdef5}
\end{align}
for all $\Lambda_1,\Lambda_2,\Lambda_3\in\Gamma(E)$ and all $f,g\in C^\infty(M)$, where $\text{Jac}$ is the Jacobiator, given by
\begin{equation} \label{eq:Jacobiator}
\text{Jac}(\Lambda_1,\Lambda_2,\Lambda_3) = [[\Lambda_1,\Lambda_2 ]\, ,\Lambda_3] + [[\Lambda_2,\Lambda_3 ]\, ,\Lambda_1] + [[\Lambda_3,\Lambda_1 ]\, ,\Lambda_2]   \, ,
\end{equation} 
and $\text{Nij}$ is the Nijenhuis operator, given by
\begin{equation} \label{eq:Nij}
\text{Nij}(\Lambda_1,\Lambda_2,\Lambda_3) =\frac{1}{6}\Big( \langle [\Lambda_1,\Lambda_2],\Lambda_3 \rangle +\langle [\Lambda_2,\Lambda_3],\Lambda_1 \rangle +\langle [\Lambda_3,\Lambda_1],\Lambda_2 \rangle \Big) \, .
\end{equation}

Let $(E,\eta,[\, ,\,]_E,\rho)$ and $(E',\eta',[\, ,\,]_{E'},\rho')$ be two Courant algebroids over the same base manifold $M$. A smooth vector bundle isomorphism $F:E\to E'$, covering the identity on $M$, is called a Courant algebroid isomorphism \cite{C-iso} if
\begin{equation} \label{eq:CA-isom}
\rho'\circ F = \rho \, , \quad \eta'\bigl(F\Lambda_1,F\Lambda_2\bigr) = \eta(\Lambda_1,\Lambda_2) \, ,\quad  F\bigl([\Lambda_1,\Lambda_2]_E\bigr) = [F\Lambda_1,F\Lambda_2]_{E'} \, ,
\end{equation}
for all $\Lambda_1,\Lambda_2\in\Gamma(E)$.

\end{document}